\documentclass[twocolumn]{jpsj3}

\usepackage{bm}
\usepackage{graphicx}
\usepackage{color}
\usepackage{afterpage}

\def\runauthor{Online-Journal Subcommittee}

\title{
A List Referring Monte-Carlo Method for Lattice Glass Models\\
}

\author{
Munetaka \textsc{Sasaki}\thanks{E-mail : msasaki@camp.apph.tohoku.ac.jp} and Koji \textsc{Hukushima}$^\dagger$
}

\inst{
Department of Applied Physics, Tohoku University, Sendai 980-8579, Japan\\
$^\dagger$Department of Basic Science, The University of Tokyo, Tokyo 153-8902, Japan
}

\recdate{\today}

\abst{
We present an effcient Monte-Carlo method for lattice glass models which are characterized 
by hard constraint conditions. The basic idea of the method is similar to that of the $N$-fold 
way method. By using a list of sites into which we can insert a particle, 
we avoid trying a useless transition which is forbidden by the constraint conditions. 
We applied the present method to a lattice glass model proposed by Biroli and M{\'e}zard. 
We first evaluated the efficiency of the method through measurements of the autocorrelation 
function of particle configurations. As a result, we found that the efficiency is much higher 
than that of the standard Monte-Carlo method. We also compared 
the efficiency of the present method with that of the $N$-fold way method in detail. 
We next examined how the efficiency of extended ensemble methods such as the replica exchange method 
and the Wang-Landau method is inflnuenced by the choice of the local update method. 
The results show that the efficiency is considerably improved by the use of efficient 
local update methods. For example, when the number of sites $N_{\rm site}$ is $1024$, 
the ergodic time $\tau_{\rm E}$ of the replica exchange method in the grand-canonical ensemble, 
which is the average round-trip time of a replica in chemical-potential space, 
with the present local update method is more than $10^2$ times shorter 
than that with the standard local update method.  This result shows that 
the efficient local update method is quite important to make extended ensemble methods more effective.}
\kword{
Monte Carlo method, $N$-fold way method, lattice glass model, hard constrained problem}

\begin{document}
\maketitle

\section{Introduction}

Lattice glass models are a kind of lattice gas model which show glassy behavior 
due to some constrained rules. The constraints are either dynamical 
rules~\cite{FredicksonAnderson84,KobAnderson93,ToninelliBiroliFisher06} 
for movements of particles or geometrical rules~\cite{BiroliMezard02,Ciamarra03,WeigtHartmann03} 
for restrictions on possible particle configurations. Lattice glass models 
have played an important role in the study of the glass. To unveil the properties of 
the actual glass, studies of lattice glass models in finite dimensions are effective. 
In such studies, numerical simulations play an important role because most of analytic 
methods such as the cavity method~\cite{MezardParisi01} are not applicable to systems 
in finite dimensions. However, it is obvious that numerical simulations in lattice glass 
models seriously suffer from the inherent characteristic of the glass, {\it i.e.}, 
slow dynamics. Because of this slow relaxation, it is very difficult to investigate 
equilibrium properties of the models. 

To relieve the problem, there are mainly two methods. The first method is 
to utilize a list which contains all of possible transitions from the current state. 
By excluding transitions which are forbidden by the constraint conditions from the list
and choosing a transition to a next state from the list, we can avoid trying a useless transition. 
The list is updated whenever the state of the system is changed. 
The idea of utilizing a list to improve the efficiency of simulation was first proposed 
by Bortz {\it et al.}~\cite{BortzKalosJebowitz75}. They proposed a rejection-free Monte-Carlo 
(MC) method in which residence time at the current state and a transition to a next state 
is chosen with proper probability by using the list. This method is called $N$-fold way method. 
It has been shown by several previous studies that $N$-fold way method is quite effective 
in lattice glass models~\cite{KobAnderson93,NakanishiTakano86}. Furthermore, an efficient MC method which utilizes a list 
has been invented for off-lattice particle systems at high densities~\cite{Mezei87}. 
The second method is to use extended ensemble methods~\cite{Iba01} 
such as the multicanonical method~\cite{BergNeuhaus91,BergNeuhaus92}, 
the Wang-Landau method~\cite{WangLandau01A,WangLandau01B}, 
and the replica exchange method~\cite{HukushimaNemoto96}. 
It is widely accepted that extended ensemble methods are effective to 
accelerate the equilibration in glassy systems.

In this study, we present an efficient MC method which utilizes a list. 
We hereafter call the method {\it list referring MC} (LRMC) method. 
This method consists of two local updates: insertion-deletion update 
and particle-hole exchange update. 
We applied the LRMC method to a lattice glass model proposed by Biroli 
and M{\'e}zard~\cite{BiroliMezard02}. We hereafter refer the model as the BM model. 
In the present study, we investigated the BM model on a regular random graph, 
which has been well examined by the cavity method~\cite{BiroliMezard02,Rivoire04,Krzakala08}, 
while the LRMC method is applicable no matter whether the model 
is defined on a sparse random graph or a usual lattice in finite dimensions. 
We evaluated the efficiency of the LRMC method through measurements of the autocorrelation function. 
As a result, we found that the relaxation time of the LRMC method is much shorter 
than that of the standard MC method, particularly at high densities. 
For example, when the chemical potential $\mu$ in the grand-canonical ensemble is 6.5, 
the relaxation time of the LRMC method is about $10^3$ times shorter than that of the 
standard MC method. 
We also compared the LRMC method with the $N$-fold way method. 
Although the particle-hole exchange update is rejection-free like the the $N$-fold way method, 
the insertion-deletion update is not rejection-free. 
Therefore, if we compare the LRMC method without the particle-hole exchange update and 
the $N$-fold way method without the particle-hole exchange update, 
the former is less efficient than the latter. However, the LRMC method with the particle-hole 
exchange update are comparable to the $N$-fold way method 
without the particle-hole exchange update in efficiency. We also found that 
the particle-hole exchange update is rather effective for the $N$-fold way method. The efficiency 
of the $N$-fold way method is considerably improved by adopting the particle-hole exchange 
update into the method.

We also examined how the efficiency of extended ensemble methods
is influenced by the choice of the local update method. 
As local update methods, we considered the LRMC and standard MC methods. 
We first investigated the influence of the local update methods on the 
the replica exchange method. In this study, we considered the grand-canonical
ensemble and performed a replica exchange simulation concerning chemical potential. 
As a result, we found that  the efficiency of the replica exchange method is greatly improved 
by the use of the LRMC method.  For example, when the number of site $N_{\rm site}$ is $1024$, 
the ergodic time $\tau_{\rm E}$, which is the average round-trip time 
of a replica in chemical-potential space, with the LRMC method is 
more than $10^2$ times shorter than that with the standard MC method.  
We next made such comparison on the Wang-Landau method. 
We measured the density of states (DOS) of the BM model by the Wang-Landau method. 
When we used the LRMC method as a local update, we succeeded in calculating the DOS
up to $N_{\rm site}=8192$. In contrast, we could not calculate the DOS even for a small size of 
$N_{\rm site}=512$ if we use the standard MC method. These results show that the efficient local update 
method is quite important to make extended ensemble methods more effective.

The outline of the paper is as follows: In \S\ref{sec:model}, we introduce the BM model. 
In \S\ref{sec:method}, we present the LRMC method. 
In \S\ref{sec:results}, we show our simulation results. 
Section \ref{sec:conclusions} is devoted to conclusions. 
Technical details for updating the list are described in Appendixes.

\section{Model}
\label{sec:model}
In this section, we introduce the BM model~\cite{BiroliMezard02} to which we apply the LRMC method. 
The BM model is a kind of lattice glass models. A binary variable $\sigma_i$ is defined on each site. 
The variable $\sigma_i$ denotes whether a site $i$ is occupied by a particle ($\sigma_i=1$) 
or not $(\sigma_i=0)$. In this study, we consider the BM model defined on a regular random graph. 
Each site is connected with $k$ neighbouring sites which are chosen randomly from all of the sites. 
A particle configuration $\{\sigma_i\}$ is restricted by hard constraints that neighbouring 
occupied sites of each particle should be less than or equal 
to $l$. The BM model is characterized by the two integers $k$ and $l$. 
They satisfy the inequality $k>l$. The probability distribution of the BM model 
for a particle configuration $\{\sigma_i\}$ is given as
\begin{equation}
P\{\sigma_i\}=Z^{-1}C\{\sigma_i\}W\{\sigma_i\}. 
\label{eqn:distribution}
\end{equation}
In this equation, $Z$ is the partition function defined by 
$Z\equiv {\rm Tr}_{\{\sigma_i\}} C\{\sigma_i\}W\{\sigma_i\}$
and $C\{\sigma_i\}$ is an indicator function which is 
one if $\{\sigma_i \}$ satisfies all of the constraint conditions or zero otherwise. 
$W\{\sigma_i\}$ is a weight of the particle configuration $\{\sigma_i\}$. 
For example, in the case of the grand-canonical ensemble, 
$W\{\sigma_i\}$ is given as
\begin{equation}
W\{\sigma_i\}=\exp\left[\mu N\{ \sigma_i \}\right],
\label{eqn:Weight_GC}
\end{equation}
where $\mu$ is a chemical potential and $N\{ \sigma_i \}$ is the number of particles 
defined by the equation
\begin{equation}
N\{\sigma_i\} \equiv \sum_{i=1}^{N_{\rm site}} \sigma_i,
\label{eqn:NofParticles}
\end{equation}
where $N_{\rm site}$ is the number of sites. 

In the present study, we will focus on the BM model on a regular
random graph with $k=3$ and $l=1$. 
All of numerical simulations are performed in this model. 
In the grand-canonical ensemble, the model exhibits a static glass transition 
with a one-step replica symmetry breaking 
at $\mu_{\rm s}\approx 6.8$~\cite{Rivoire04,Krzakala08,HukushimaSasa10}. 
The close-packing density of the model is estimated to be $0.57574$ 
by the cavity method~\cite{Rivoire04,Krzakala08}.

\section{LRMC Method}
\label{sec:method}
\subsection{Standard MC method and its drawbacks}
In this subsection, we explain a standard MC method and its drawbacks. 
The following is the procedure of the standard MC method with the Metropolis transition 
probability~\cite{Metropolis53}: 
\begin{itemize}
\item[(a)] Prepare an initial state. 
The initial state can be chosen arbitrarily if it satisfies the constraint conditions. 
\item[(b)] Choose a site $k$ at random. 
\item[(c)] Create a new state $\{ \sigma_i' \}$ from the current state $\{ \sigma_i \}$ 
by changing the value of $\sigma_k$ from $0$ to $1$ or vice versa. 
\item[(d)] Accept the change into $\{ \sigma_i' \}$ with the probability
\begin{equation}
A(\{ \sigma_i \} \rightarrow \{ \sigma_i' \})
= C\{ \sigma_i' \} \min \left( 1, \frac{W\{ \sigma_i' \}}{W\{ \sigma_i \}} \right).
\label{eqn:AP_standard}
\end{equation}
If it is not accepted, unchange the state from $\{ \sigma_i \}$.
\item[(e)] Return to (b) and repeat the steps (b)-(d). 
\end{itemize}

We next explain the drawback of this standard MC method. We hereafter consider the grand-canonical 
ensemble whose equilibrium weight is given by Eq.~(\ref{eqn:Weight_GC}). Then, the acceptance ratio 
in Eq.~(\ref{eqn:AP_standard}) is rewritten as 
\begin{equation}
A(\{ \sigma_i \} \rightarrow \{ \sigma_i' \}) =\left\{
\begin{array}{cc}
C\{ \sigma_i' \} & (\sigma_k=0),\vspace{2mm}\\
\exp(-\mu) & (\sigma_k=1),\\
\end{array}
\right.
\end{equation}
where we have used Eq.~(\ref{eqn:Weight_GC}) and assumed that $\mu$ is not negative 
for simplicity. 
We also have used the fact that $C\{ \sigma_i' \}$ 
is $1$ when $\sigma_k=1$ because deletion of a particle 
never conflicts with the constraint conditions.
When $\mu$ is large, there are few empty sites at which we can insert a particle. 
Therefore, when $\sigma_k=0$, the trial to insert a particle into the site $k$ fails 
in most cases. On the other hand, when $\sigma_k=1$, the acceptance ratio 
$A(\{ \sigma_i \} \rightarrow \{ \sigma_i' \})$ is quite small 
because $\mu$ is large. As a result of the small acceptance ratio in step (d), 
relaxation to the equilibrium state becomes very slow. 
In order to overcome the difficulty in the standard MC method, 
we introduce two efficient local updates for the LRMC method 
in the following two subsections.

\subsection{Local update I: insertion-deletion update}
\label{subsec:LocalUpdate1}

In this subsection, we explain the first update which consists 
of insertion and deletion of a particle. We hereafter call it {\it insertion-deletion update}. 
The basic idea is as follows: 
Because it is a waste of computational time trying insertion of 
a particle which is forbidden by the constraint conditions, we just try insertion which does not 
conflict with the constraint conditions. In order to do that, we make a list of the sites 
into which we can insert a particle and update the list whenever the particle configuration is changed. 
We choose an insertion site at random from the list. The acceptance ratio of the 
insertion and that of the deletion are chosen so that the detailed balance condition is satisfied. 
We determine whether we try insertion or deletion of a particle with the equal probability. 

As mentioned before, this method is rather similar to that of the $N$-fold way 
method~\cite{BortzKalosJebowitz75} in the sense that we utilize a list 
to improve the simulation efficiency. It is worth pointing out that 
we have to pay some computational cost to make and update the list. 
It will be discussed in detail in \S\ref{subsec:efficiency} whether the LRMC method is 
still effective or not even if this additional computational cost is taken into account.

We now start concrete description of the insertion-deletion update. 
We assume that we have a list of the sites 
from which we can delete a particle and that of the sites into which we can insert a particle. 
The former list is the same as that of the occupied sites because deletion of a particle 
never conflicts with the constraint conditions. The method to detect the sites on which 
the insertion list has to be updated and the method to update the insertion list are explained 
in appendices~\ref{sec:appendixA} and~\ref{sec:appendixB}, respectively.
The following is the flow chart of the insertion-deletion update:
\begin{itemize}
\item[(1)] Choose whether we try insertion or deletion with the equal probability. 
\item[(2a)] If the insertion is chosen in step (1), select an insertion site at random from the 
insertion list and accept the insertion with an acceptance ratio 
\begin{equation}
A_{\rm I}(\{\sigma_i\}\rightarrow \{\sigma_i'\}) = 
\min \left(1,\frac{K\{\sigma_i\} W\{\sigma_i'\}}
{N\{\sigma_i'\} W\{\sigma_i\}}\right),
\label{eqn:AcceptanceRatio_I}
\end{equation}
where $K\{\sigma_i\}$ is the number of the sites into which we can insert a particle, 
$N\{\sigma_i\}$ is the number of particles defined by Eq.~(\ref{eqn:NofParticles}), 
and $\{\sigma_i'\}$ is the particle configuration created from $\{\sigma_i\}$ 
by inserting a particle into the insertion site. 
\item[(2b)] If the deletion is chosen, select a deletion site at random from the deletion list 
and accept the deletion with an acceptance ratio 
\begin{equation}
A_{\rm R}(\{\sigma_i\}\rightarrow \{\sigma_i'\}) = 
\min \left(1,\frac{N\{\sigma_i\} W\{\sigma_i'\}}
{K\{\sigma_i'\} W\{\sigma_i\}}\right),
\label{eqn:AcceptanceRatio_R}
\end{equation}
where $\{\sigma_i'\}$ is the particle configuration created from $\{\sigma_i\}$ 
by deleting a particle from the deletion site. 
\item[(3)] Update the insertion and deletion lists when the insertion or deletion is accepted. 
\item[(4)] Return to (1) and repeat the steps (1)-(3). 
\end{itemize}

It is straightforward to show that the procedure described above
satisfies the detailed balance condition. We consider two particle configurations 
$\{ \sigma_i \}$ and $\{ \sigma_i' \}$. The latter configuration $\{ \sigma_i' \}$ is created 
from $\{ \sigma_i \}$ by inserting a particle into an insertion site. The transition probability 
$T(\{ \sigma_i \}\rightarrow \{ \sigma_i' \})$ from $\{ \sigma_i \}$ to $\{ \sigma_i' \}$ 
is given as
\begin{eqnarray}
T(\{ \sigma_i \}\rightarrow \{ \sigma_i' \}) 
= \frac{1}{2} \times \frac{1}{K\{ \sigma_i \}} \times 
A_{\rm I}(\{ \sigma_i \}\rightarrow \{ \sigma_i' \}).
\label{eqn:LU1_TPA}
\end{eqnarray}
In the right hand side of Eq.~(\ref{eqn:LU1_TPA}), 
the first factor is the probability that the insertion is chosen in step (1), the second factor 
is the probability that the proper insertion site is chosen 
among the $K\{\sigma_i \}$ insertion sites in step (2a), and the third factor is the probability that 
the transition into $\{\sigma_i'\}$ is accepted in step (3). Similarly, the transition probability 
of the reversal process $T(\{ \sigma_i' \}\rightarrow \{ \sigma_i \})$ is given as
\begin{eqnarray}
T(\{ \sigma_i' \}\rightarrow \{ \sigma_i \}) 
= \frac{1}{2} \times \frac{1}{N\{ \sigma_i' \}} \times 
A_{\rm R}(\{ \sigma_i' \}\rightarrow \{ \sigma_i \}). 
\label{eqn:LU1_TPB}
\end{eqnarray}
The second factor in the right hand side of Eq.~(\ref{eqn:LU1_TPB}) comes from the fact that 
the proper deletion site is chosen from the $N\{ \sigma_i' \}$ occupied sites. The detailed balance condition
\begin{equation}
P\{\sigma_i\}T(\{ \sigma_i' \}\rightarrow \{ \sigma_i \})
=P\{\sigma_i'\}T(\{ \sigma_i \}\rightarrow \{ \sigma_i' \}),
\label{eqn:DBcondition}
\end{equation}
can be shown from Eqs.~(\ref{eqn:distribution}), (\ref{eqn:AcceptanceRatio_I}), (\ref{eqn:AcceptanceRatio_R}), 
(\ref{eqn:LU1_TPA}), (\ref{eqn:LU1_TPB}), and the fact that $C\{\sigma_i\}=C\{\sigma_i'\}=1$.

\subsection{Local update II: particle-hole exchange update}
\label{subsec:LocalUpdate2}
In this subsection, we explain the second local update which consists of exchanges 
of a particle for a hole. The basic idea is similar to that of spin-exchange 
method~\cite{Kawasaki66,KawasakiBook}. 
We hereafter call it {\it particle-hole exchange update}. 
The acceptance ratio of this local update is unity 
when the weight $W\{ \sigma_i \}$ depends only on the number of particles. 
The grand-canonical ensemble given as Eq.~(\ref{eqn:Weight_GC}) is an example 
which satisfies this condition. We hereafter consider the case that this condition is 
satisfied. As in the previous subsection, it is assumed that we have the insertion and 
deletion lists. The flow chart of the particle-hole exchange update is as follows:
\begin{itemize}
\item[(1)] Select a deletion site at random from the deletable sites. 
Then, delete a particle from the site. 
\item[(2)] Update the insertion and deletion lists. 
\item[(3)] Select an insertion site at random 
from the insertable sites except the deletion site in step (1). 
Then, insert a particle into the site. 
\item[(4)] Update the insertion and deletion lists. 
\item[(5)] Return to (1) and repeat the steps (1)-(4). 
\end{itemize}
In the step (3), it is forbidden that the deletion site in step (1) is chosen 
as the insertion site so that this exchange process causes a change 
in the particle configuration. However, in practice, this useless 
exchange is allowed if the deletion site in step (1) is the only insertable site.

\begin{figure}[t]
\begin{center}
\includegraphics[width=\columnwidth]{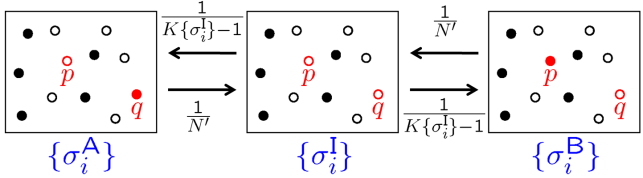}
\end{center}
\caption{(Color online) 
Transition processes between $\{ \sigma_i^{\rm A} \}$ and $\{ \sigma_i^{\rm B} \}$ 
in the particle-hole exchange update. $\{ \sigma_i^{\rm I} \}$ is an intermediate state 
(see text for its definition). The transition probabilities among the states are 
shown above and below the arrows. 
}
\label{fig:Pexchange}
\end{figure}

We now show that this procedure satisfies the detailed balance condition. 
We consider two particle configurations $\{ \sigma_i^{\rm A} \}$ and $\{ \sigma_i^{\rm B} \}$ 
that are transferred from each other by a particle-hole exchange update, 
as shown in Fig.~\ref{fig:Pexchange}. 
The two configurations are the same except at the two sites $p$ and $q$. 
We assume that the number of particles in the two configurations is $N'$. 
In Fig.~\ref{fig:Pexchange}, $\{ \sigma_i^{\rm I} \}$ is an intermediate state which is 
realized after the step (1) of the exchange process. We see that the two sites 
$p$ and $q$ are empty in the intermediate state. It is important to notice that 
the intermediate state in the transition process from $\{ \sigma_i^{\rm A} \}$ to 
$\{ \sigma_i^{\rm B} \}$ is the same as that in the reverse process. 
As shown in Fig.~\ref{fig:Pexchange}, the transition from $\{ \sigma_i^{\rm A} \}$ 
to $\{ \sigma_i^{\rm B} \}$ occurs if and only if the site $q$ is 
chosen from the $N'$ deletable sites in step (1) and the site $p$ is chosen 
from the $K\{\sigma_i^{\rm I} \}-1$ insertable sites in step (3) 
(recall that the deletion site in step (1) is not chosen as the insertion site). 
Therefore, the transition probability $T(\{ \sigma_i^{\rm A} \} \rightarrow \{ \sigma_i^{\rm B} \})$ 
is $\{N'(K\{\sigma_i^{\rm I} \}-1)\}^{-1}$. The transition probability of the reverse process 
$T(\{ \sigma_i^{\rm B} \} \rightarrow \{ \sigma_i^{\rm A} \})$ is also 
$\{N'(K\{\sigma_i^{\rm I} \}-1)\}^{-1}$ in the same way. As a result, we find
\begin{equation}
T(\{ \sigma_i^{\rm A} \} \rightarrow \{ \sigma_i^{\rm B} \}) = 
T(\{ \sigma_i^{\rm B} \} \rightarrow \{ \sigma_i^{\rm A} \}).
\label{eqn:DBCprove_EX_A}
\end{equation}
On the other hand, when the weight of each state depends only on the number of particles, 
we obtain 
\begin{equation}
P\{ \sigma_i^{\rm A} \}=P\{ \sigma_i^{\rm B} \},
\label{eqn:DBCprove_EX_B}
\end{equation}
because $C\{ \sigma_i^{\rm A} \}=C\{ \sigma_i^{\rm B} \}=1$ and the number of particles is the same 
in the two configurations. It is clear from Eqs.~(\ref{eqn:DBCprove_EX_A}) and 
(\ref{eqn:DBCprove_EX_B}) that the detail balance condition Eq.~(\ref{eqn:DBcondition}) is 
satisfied. 

It should be noted that the sampling from the distribution Eq.~(\ref{eqn:distribution}) 
can not be achieved by only repeating the exchange process because it preserves 
the number of particles. It is necessary to combine the insertion-deletion update 
with the particle-hole exchange update to yield the sampling 
from the distribution Eq.~(\ref{eqn:distribution}).

\section{Results}
\label{sec:results}
This section is devoted to show our simulation results. 
For comparison, simulation is performed not only by the LRMC method 
but also by the standard MC method. In standard MC simulations, 
we adopted the Metropolis transition probability. 
All of simulations are performed for the BM model on regular random graphs 
with $k=3$ and $l=1$.

\begin{figure}[t]
\begin{center}
\includegraphics[width=0.8\columnwidth]{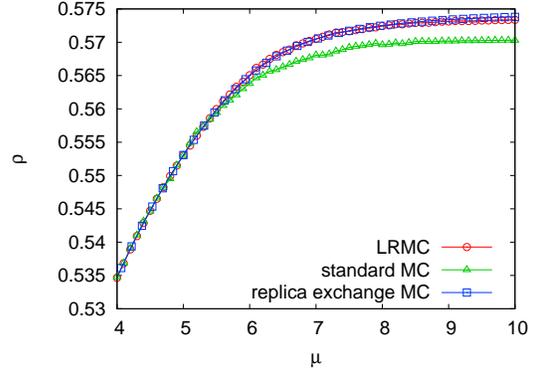}
\end{center}
\caption{(Color online) The chemical potential $\mu$ dependences 
of the average occupation density $\rho(\mu)$ measured by three different 
simulation methods are compared. The data obtained by the LRMC method, standard MC method, 
and replica exchange method are denoted by open circles, open triangles, 
and open squares, respectively. The model is the BM model on regular random graphs 
with $k=3$ and $l=1$. The number of sites $N_{\rm site}$ is $512$. 
The average over random graphs is taken over $100$ samples.  
}
\label{fig:AveDensity}
\end{figure}

\subsection{The chemical-potential dependence of the average occupation density}
\label{subsec:AveDensity}
To confirm that correct results are obtained by the LRMC method, 
we first measured the average occupation density $\rho$ 
in the grand-canonical ensemble by both the LRMC and standard MC methods 
and compare their results. We adopted a simulated annealing method: 
The chemical potential $\mu$ is gradually increased from $0$ to $10$ 
in steps of $\Delta \mu=0.01$. The system is kept at each chemical 
potential for $10^5$ MC steps, where one MC step of the LRMC method is defined by $N_{\rm site}$ 
trials of the insertion-deletion update and subsequent $N_{\rm site}$ particle-hole exchange 
updates. The first $5\times 10^4$ MC steps are for relaxation and the subsequent 
$5\times 10^4$ MC steps are for measurement. The number of sites $N_{\rm site}$ is $512$.
The average over random graphs is taken over $100$ samples.

The results of measurements of the average occupation density $\rho(\mu)$ are shown 
in Fig.~\ref{fig:AveDensity}. When $\mu$ is small, both the data 
coincide with each other. This shows that correct results are obtained 
by the LRMC method. On the other hand, for larger $\mu$, 
$\rho(\mu)$ obtained by the LRMC method is clearly larger than 
that obtained by the standard MC method. 
In Fig.~\ref{fig:AveDensity}, we also show the data obtained 
by the replica exchange method. The data of the LRMC method are slightly 
smaller than those of the replica exchange method for large $\mu$. 
However, their difference is rather small. 
These facts indicate that the equilibration is accelerated by the use of the LRMC method.

\subsection{The size and chemical-potential dependence of the acceptance ratio 
of the insertion-deletion update}
\label{subsec:AcceptanceRatio}

We next measured the size and chemical-potential dependence of the acceptance ratio 
$P_{\rm accept}$ in the grand-canonical ensemble. The measurement was performed 
in both the LRMC and standard MC methods for comparison. 
Because the acceptance ratio of the particle-hole exchange update is unity, 
we measured the acceptance ratio only for the insertion-deletion update. 
The conditions of measurements are the same as those of average occupation 
density $\rho$ in \S\ref{subsec:AveDensity}. 
During the simulation, we also measured the average of $K\{\sigma_i\}$ at each $\mu$.

\begin{figure}[t]
\begin{center}
\includegraphics[width=0.8\columnwidth]{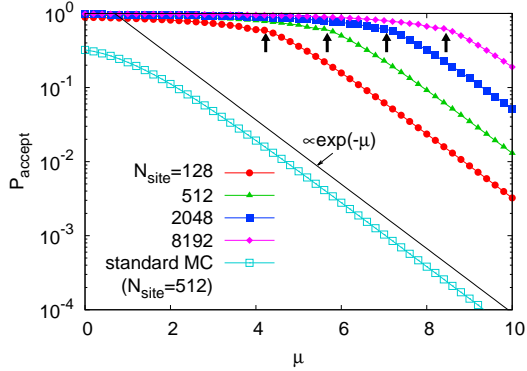}
\end{center}
\caption{(Color online) The chemical potential $\mu$ dependence of the acceptance ratio $P_{\rm accept}$
for the BM model on regular random graphs with $k=3$ and $l=1$. 
The data measured by the LRMC method are denoted by 
full symbols and those measured by the standard MC method are denoted 
by open squares. The number of sites $N_{\rm site}$ in the standard MC method 
is $512$. The four bold arrows in the figure indicate the value of 
$\mu^*(N_{\rm site})$ defined by Eq.~(\ref{eqn:def_mu*}) at which 
the average number of the insertable sites is one. 
The solid line is proportional to $\exp(-\mu)$.}
\label{fig:PacceptB}
\end{figure}

In Fig.~\ref{fig:PacceptB}, the acceptance ratio $P_{\rm accept}$ is plotted 
as a function of $\mu$. The data measured 
by the LRMC method are denoted by full symbols and those measured 
by the standard MC method are denoted by open squares. We find that 
the acceptance ratio of the LRMC method is two or three orders of magnitude higher 
than that of the standard MC method when $\mu$ is large. 
We also see that the acceptance ratio of the LRMC method 
depends on not only the chemical potential $\mu$ but also 
the size $N_{\rm site}$. For a fixed $\mu$, it increases with the size. 
In contrast, we have checked that the acceptance ratio of the standard MC method 
hardly depends on the size (not shown in the figure).

\begin{figure}[t]
\begin{center}
\includegraphics[width=0.8\columnwidth]{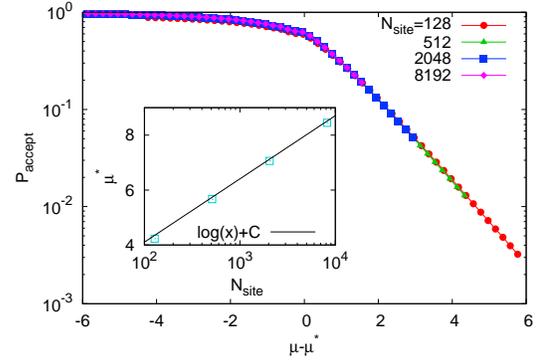}
\end{center}
\caption{(Color online) Scaling plot of the acceptance ratio $P_{\rm accept}$ 
of the LRMC method. In the figure, $P_{\rm accept}$'s  of the LRMC method 
in Fig.~\ref{fig:PacceptB} (full symbols) are plotted as a function of $\mu-\mu^*$. 
The inset shows the size dependence of $\mu^*$ which is defined 
by Eq.~(\ref{eqn:def_mu*}). 
}
\label{fig:PacceptC}
\end{figure}

The four arrows in the figure indicate the value of chemical 
potential $\mu^*(N_{\rm site})$ at which the average number of the insertable 
sites is unity. To be specific, $\mu^*(N_{\rm site})$ is defined by 
\begin{equation}
\overline{\langle K\{\sigma_i \} \rangle}_{N_{\rm site}, \mu^*(N_{\rm site})}=1.
\label{eqn:def_mu*}
\end{equation}
In this equation, the overline $\overline{\cdots}$ 
denotes the average over different realizations of regular random graphs and 
the bracket $\langle \cdots \rangle$ denotes the average 
in the grand-canonical ensemble. The two subscripts of the bracket denote 
the size and chemical potential. We see from Fig.~\ref{fig:PacceptB} 
that $P_{\rm accept}$ decays exponentially above $\mu^*(N_{\rm site})$. 
We also notice that $P_{\rm accept}$ is mostly determined by the difference 
$\mu-\mu^*(N_{\rm site})$. In Fig.~\ref{fig:PacceptC}, $P_{\rm accept}$ 
is plotted as a function of $\mu-\mu^*(N_{\rm site})$. All of the data 
nicely collapse into a single curve. The inset of Fig.~\ref{fig:PacceptC}
shows the size dependence of $\mu^*(N_{\rm site})$. 
We see that $\mu^*(N_{\rm site})$ is approximately given as
\begin{equation}
\mu^*(N_{\rm site}) \approx \log(N_{\rm site})+C.
\label{eqn:Sdependence_mu*}
\end{equation}

Now let us consider how this behavior of acceptance ratio is understood. 
It is naturally expected that the acceptance ratio is close to unity when $\mu\le\mu^*$ 
because there is at least one insertable site in this case. We can also understand 
this behavior from a relation between $N\{\sigma_i\}$ and $K\{\sigma_i\}$.
As shown in appendix~\ref{sec:NewAppendix}, the average number of insertable sites 
$\langle K\{\sigma_i\} \rangle_{\rm \mu}$ in the ground-canonical ensemble 
is related to that of particles $\langle N\{\sigma_i\} \rangle_{\rm \mu}$ by 
\begin{equation}
\langle K\{ \sigma_i \} \rangle_{\mu}={\rm e}^{-\mu}\langle N\{ \sigma_i \} \rangle_{\mu}.
\label{eqn:KNrelation_final}
\end{equation}
We emphasize that this is a static relation and it is valid  for any Monte-Carlo 
methods which realize the grand-canonical ensemble defined by Eq.~(\ref{eqn:Weight_GC}). 
From Eqs.~(\ref{eqn:Weight_GC}) and (\ref{eqn:KNrelation_final}), we find that 
the two acceptance ratios given by Eqs.~(\ref{eqn:AcceptanceRatio_I}) and 
(\ref{eqn:AcceptanceRatio_R}) become unity if the number of particles $N\{\sigma_i\}$ and 
that of the insertable sites $K\{\sigma_i\}$ are equal to their mean values.

We next consider the size dependence of $\mu^*$. From Eqs.~(\ref{eqn:def_mu*}) and (\ref{eqn:KNrelation_final}), 
we obtain
\begin{equation}
\overline{\langle K\{ \sigma_i \} \rangle}_{N_{\rm site}, \mu^*}=
\rho(N_{\rm site},\mu^*)N_{\rm site}{\rm e}^{-\mu^*}=1, 
\label{eqn:mustar_first}
\end{equation}
where $\rho\equiv \overline{\langle N\{\sigma_i\}/N_{\rm site} \rangle}$ 
is the average occupation density. Because $\rho$ hardly depends on 
the size and its $\mu$ dependence is much weaker than ${\rm e}^{-\mu}$ 
(see Fig.~\ref{fig:AveDensity}), it is appropriate 
to approximate $\rho$ by a constant, leading to Eq.~(\ref{eqn:Sdependence_mu*}). 

Lastly, we consider how the exponential decay of $P_{\rm accept}$ above $\mu^*(N_{\rm site})$ 
is understood. We first focus on the acceptance ratio for deletion. 
When $\mu > \mu^*(N_{\rm site})$, $K\{\sigma_i'\}$ in Eq.~(\ref{eqn:AcceptanceRatio_R}) 
is very small. However, it is always larger than one because $\{ \sigma_i' \}$ is 
a particle configuration {\it after} a particle is removed from a site. 
Note that it is always possible to insert a particle into the deletion site. 
Therefore, when $\mu \gg \mu^*$, $K\{\sigma_i'\}$ in Eq.~(\ref{eqn:AcceptanceRatio_R}) 
is well approximated by one. Then, by regarding $N\{\sigma_i\}$ 
in Eq.~(\ref{eqn:AcceptanceRatio_R}) as a constant and using Eq.~(\ref{eqn:Weight_GC}), 
we obtain
\begin{equation}
A_{\rm R}(\{\sigma_i\}\rightarrow \{\sigma_i'\}) \propto {\rm e}^{-\mu}.
\label{eqn:Edecay_3rd}
\end{equation}
We next turn to the acceptance ratio for the insertion. 
Now the point is that the average number of particles does not change 
once the system is equilibrated. Therefore, because the insertion process and deletion process 
are chosen with the equal probability, the equilibrium value of $A_{\rm R}$ and that of 
$A_{\rm I}$ should be the same. This means that $A_{\rm I}$ is also given by Eq.~(\ref{eqn:Edecay_3rd}). 
These are the reasons why $P_{\rm accept}$ in Fig.~\ref{fig:PacceptB} decays exponentially 
when $\mu > \mu^*$. On the other hand, as we discussed above, the acceptance ratio 
$P_{\rm accept}$ is close to unity when $\mu < \mu^*$. 
Therefore, one can naturally expect that $P_{\rm accept}$ satisfies a scaling law
\begin{equation}
P_{\rm accept}(N_{\rm site},\mu) = G[\mu-\mu^*(N_{\rm site})],
\end{equation}
where $G$ is a scaling function which behaves as
\begin{equation}
G(X)=\left\{
\begin{array}{cc}
1 & (X \ll 0), \\
{\rm e}^{-X} & (X \gg 0).
\end{array}
\right.
\end{equation}
As mentioned above, the validity of the scaling is nicely demonstrated 
in Fig.~\ref{fig:PacceptC}.

\subsection{Estimation of the efficiency}
\label{subsec:efficiency}
In order to estimate the efficiency of the LRMC method quantitatively, 
we measure the autocorrelation function of particle configurations defined by 
\begin{eqnarray}
C(t)=\frac{\sum_i \overline{\langle \sigma_i(t+t')\sigma_i(t')\rangle}
-\sum_i \overline{\langle \sigma_i(t')\rangle^2}}
{\sum_i \overline{\langle \sigma_i(t')\sigma_i(t')\rangle}
-\sum_i \overline{\langle \sigma_i(t')\rangle^2}}. 
\label{eqn:DefAF}
\end{eqnarray}
The average over random graphs was taken 
over $100$ samples and the average in the grand-canonical ensemble was taken over $320$ MC 
runs with different random number sequences. Therefore, we performed $32000$ MC runs to calculate 
$C(t)$ for each $\mu$. The time for equilibration, {\it i.e.}, $t'$ in Eq.~(\ref{eqn:DefAF}), 
is chosen to be sufficiently larger than the relaxation time of the autocorrelation function.

\begin{figure}[t]
(i)
\begin{center}
\includegraphics[width=0.8\columnwidth]{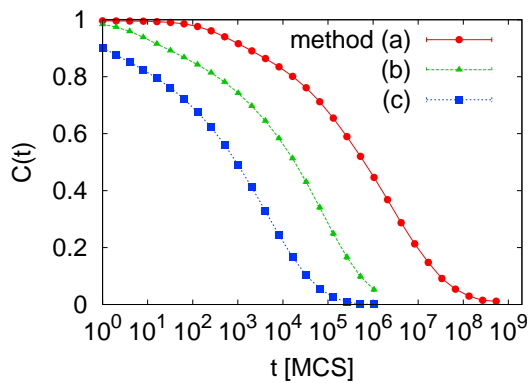}
\end{center}
(ii)
\begin{center}
\includegraphics[width=0.8\columnwidth]{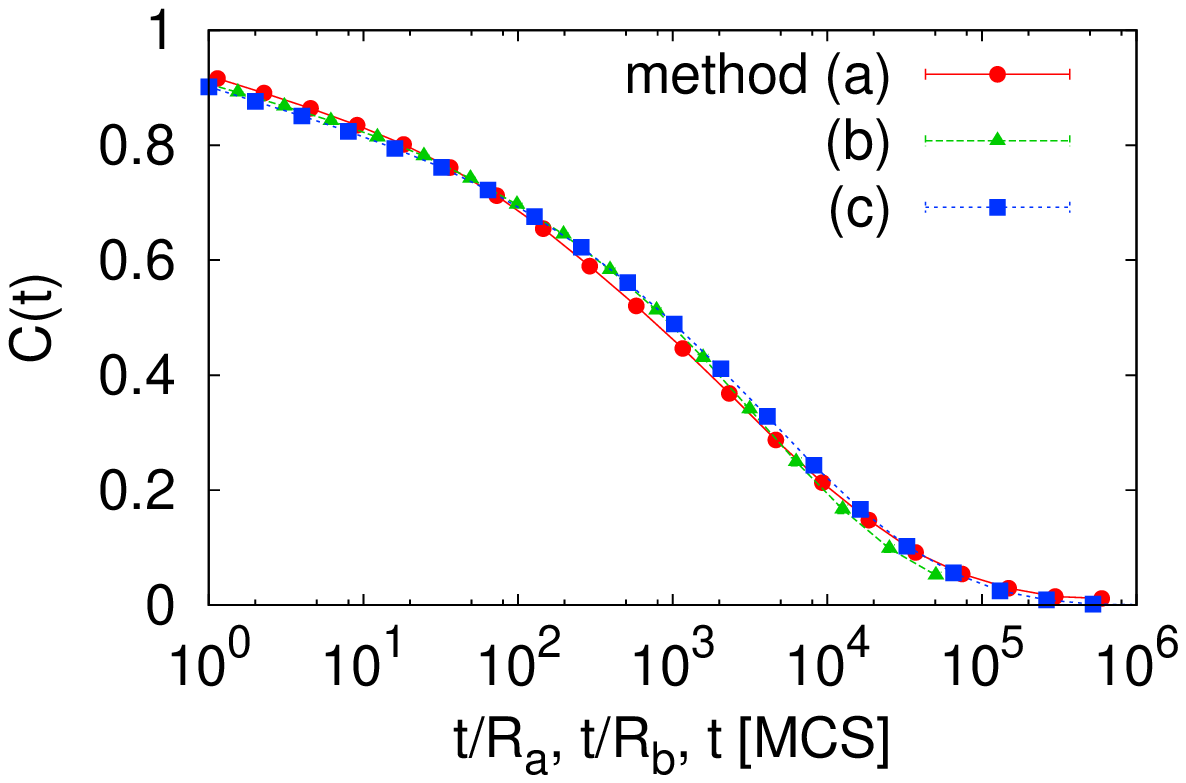}
\end{center}
\caption{(Color online) (i) The time dependence of the autocorrelation functions $C(t)$'s measured 
with the three different simulation methods and (ii) their scaling plot. 
The model is the BM model on regular random graphs with $k=3$ and $l=1$. 
The number of sites $N_{\rm site}$ is $512$ and 
the chemical potential $\mu$ is $6.5$.  The average over random graphs is taken 
over $100$ samples. For each sample, the average over thermal noise is taken over 
$320$ independent MC runs with different random number sequences. In Fig.~(ii), $C(t)$'s measured 
with the three simulation methods (a), (b), and (c) are plotted 
as a function of $t/R_{\rm a}$, $t/R_{\rm b}$, and $t$, respectively. $R_{\rm a}$ and $R_{\rm b}$ are 
evaluated to be $901$ and $20.8$, respectively, by fitting. 
}
\label{fig:Corr_3methods}
\end{figure}

To make comparisons between the LRMC and standard MC methods,
we performed MC simulations with the following three different methods:
\begin{itemize}
\item[(a)] Standard MC method. 
\item[(b)] LRMC method with only the insertion-deletion update. 
\item[(c)] LRMC method with both the insertion-deletion and particle-hole exchange updates. 
\end{itemize}
The autocorrelation function defined by Eq.~(\ref{eqn:DefAF}) is measured in each of 
the three simulation methods to compare their efficiencies. 
In the simulation method (b), one MC step is defined by $N_{\rm site}$ trials of 
the insertion-deletion update. 
In the method (c), one MC step is defined by $N_{\rm site}$ trials of 
the insertion-deletion update and subsequent $N_{\rm site}$ particle-hole 
exchange updates. The method (c) was used in the measurements of the average occupation 
density (Fig.~\ref{fig:AveDensity}) and the acceptance ratio (Figs.~\ref{fig:PacceptB} 
and~\ref{fig:PacceptC}).

\begin{figure}[t]
\begin{center}
\includegraphics[width=0.8\columnwidth]{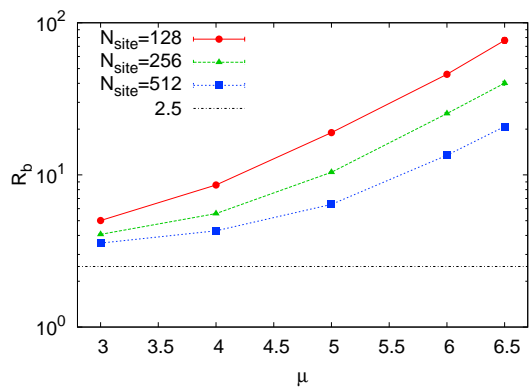}
\end{center}
\caption{(Color online) The $\mu$ dependence of the scaling factor 
$R_{\rm b}$ for three sizes. 
The model is the BM model on regular random graphs with $k=3$ and $l=1$. 
The particle-hole exchange update is effective if $R_{\rm b}$ 
is larger than a threshold value $2.5$ depicted by the dash-dotted line (see text for details).
}
\label{fig:Rb}
\end{figure}

In Fig.~\ref{fig:Corr_3methods}~(i), we show $C(t)$'s measured with the three simulation methods 
as a function of MC steps. The number of sites $N_{\rm site}$ is $512$ and 
the chemical potential $\mu$ is $6.5$.  We see that $C(t)$'s of the methods (b) and (c) 
decay much faster than that of the method (a). We next try a scaling of these data. 
The result is shown in Fig.~\ref{fig:Corr_3methods}~(ii). 
In the figure, $C(t)$ of the method (c) is plotted as a function of $t$, 
whereas $C(t)$ of the method (a) and that of the method (b) are plotted as a function of 
$t/R_{\rm a}$ and $t/R_{\rm b}$, respectively. The scaling factors $R_{\rm a}$ and $R_{\rm b}$ 
are evaluated by fitting. We see that all of the data collapse into 
a single curve. We confirmed that this scaling holds well for all of the sizes 
(from $128$ to $512$) and chemical potentials (from $0$ to $6.5$) we examined. 
These results indicate that the local updates introduced in the LRMC method 
do not change intrinsic dynamics of the system. 
By doing such analyses, we evaluated $R_{\rm a}$ and $R_{\rm b}$ for several 
$N_{\rm site}$'s and $\mu$'s. 

We see from Fig.~\ref{fig:Corr_3methods}~(i) that $C(t)$ of the method (c) decays 
faster than that of the method (b). However, because the computational time of 
the method (c) per one MC step is larger than that of the method (b), 
it is not clear solely from this result whether the particle-hole 
exchange update is really effective or not. 
Therefore, we first measured the computational times of the two methods 
per one MC step. As a result, we found that the computational time of 
the method (c) is about $2.5$ times larger than that of the method (b) 
regardless of the size and chemical potential. This means that 
the particle-hole exchange update is effective if the scaling factor $R_{\rm b}$ is larger than $2.5$. 

\begin{figure}[t]
\begin{center}
\includegraphics[width=0.8\columnwidth]{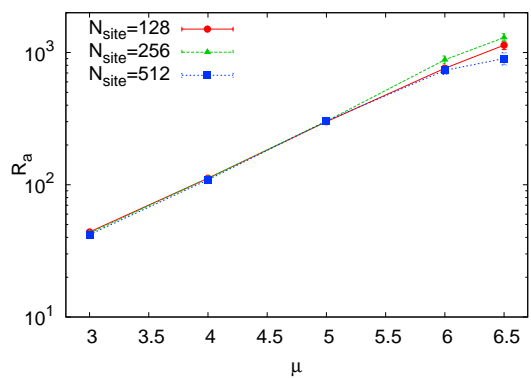}
\end{center}
\caption{(Color online) 
The $\mu$ dependence of the scaling factor $R_{\rm a}$ for three sizes. 
The model is the BM model on regular random graphs with $k=3$ and $l=1$. 
}
\label{fig:Ra}
\end{figure}

Figure~\ref{fig:Rb} shows how $R_{\rm b}$ depends on the size and chemical potential. 
We notice that $R_{\rm b}$ is larger than $2.5$ for all of the sizes and chemical potentials 
we examined. This shows that the particle-hole exchange update 
is always effective regardless of 
$N_{\rm site}$ and $\mu$. We also find that 
\begin{itemize}
\item [1)] $R_{\rm b}$ increases with $\mu$. 
\item [2)] $R_{\rm b}$ increases with decreasing $N_{\rm site}$. 
\end{itemize}
The fact 1) shows that the particle-hole exchange update is more effective when $\mu$ is large. 
The fact 2) indicates that the relative importance of the particle-hole exchange update 
increases with decreasing the size because the acceptance ratio of the insertion-deletion update 
decreases with decreasing the size (see Fig.~\ref{fig:PacceptB}) 
and that of the particle-hole exchange update 
is always unity.

Now let us turn to the comparison between the LRMC and standard MC methods. 
Because the method (c) is always more efficient than the method (b), it is enough 
to compare the two methods (a) and (c). Then, we first evaluated
the scaling factor $R_{\rm a}$ for several sizes and chemical potentials. 
Figure~\ref{fig:Ra} shows the results. We see that the ratio increases exponentially 
with $\mu$. This means that the superiority of the LRMC method to the standard MC method 
increases rapidly with $\mu$. To understand this behavior of $R_{\rm a}$, it is worth 
recalling that the acceptance ratio of the standard MC method decays exponentially
(see Fig.~\ref{fig:PacceptB}). On the other hand, such rapid decrease 
in the acceptance ratio does not exist in the LRMC method. 
As shown in Fig.~\ref{fig:PacceptB}, the acceptance ratio of the insertion-deletion update 
decays more gradually than that of the standard MC method. Furthermore, 
the acceptance ratio of the particle-hole exchange update is unity. This difference in the 
acceptance ratio is probably the main reason why the LRMC method is 
much more efficient than the standard MC method when $\mu$ is large.

We next compared the two methods from a view point of computational time. As a result, 
we found that, if the standard MC method is implemented in a usual way, 
the computational time of the LRMC method per one MC step 
is about $7$ times larger than that of the standard MC method. 
The efficiency of the LRMC method is estimated by the number $R_{\rm a}$ divided 
by this ratio of the computational time. 
Therefore, Fig.~\ref{fig:Ra} shows that the LRMC method is much more efficient than 
the standard MC method for large $\mu$ 
even given the smallness of the computational time of the standard MC method. 
However, we can reduce the computational time of the standard MC method greatly 
by using a multi-spin coding technique~\cite{BhanotDukeSalvador86A,BhanotDukeSalvador86B}, 
which is a special numerical method for models with discrete, 
especially binary, variables. 
Unfortunately, this technique is not applicable to the LRMC method. 
We found that, when this technique is used, the computational time 
of the LRMC method is about $230$ times larger than 
that of the standard MC method. This means that the superiority of the LRMC 
method might be reduced considerably by the use of the multi-spin coding technique. 
However, when $\mu=6.5$, the LRMC method is still about 
$5$ times more efficient than the standard MC method. We also remark that the multi-spin coding technique 
is not always applicable when it is used with an extended ensemble method. 
For instance, the multi-spin coding technique is not incompatible with the Wang-Landau method that is known 
to be efficient for evaluating the density of states, because the weight $W\{\sigma_i\}$ in 
Eq.~(\ref{eqn:Weight_GC}) always changes during the simulation. Meanwhile, as demonstrated in \S\ref{subsec:DOS}, 
the LRMC method can be efficiently coupled with the Wang-Landau method. 

\begin{figure}[t]
\begin{center}
\includegraphics[width=0.8\columnwidth]{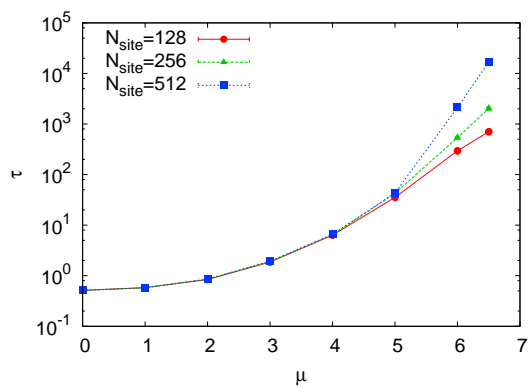}
\end{center}
\caption{(Color online) The $\mu$ dependence of the relaxation time $\tau$ 
of the method (c) for three sizes. The model is 
the BM model on regular random graphs with $k=3$ and $l=1$. 
$\tau$ is calculated by Eq.~(\ref{eqn:RelaxationTime_cal}).} 
\label{fig:RelaxationTime}
\end{figure}

Lastly, we examined how the relaxation time depends on the size and chemical potential. 
The result is shown in Fig.~\ref{fig:RelaxationTime}. Particle configuration 
$\{\sigma_i\}$ is updated with the method (c) and the relaxation time $\tau$ is 
defined by the integral
\begin{equation}
\tau\equiv \int_0^{t_{\rm max}} {\rm d}t'C(t'), 
\label{eqn:RelaxationTime_cal}
\end{equation}
where $t_{\rm max}$ is the maximum time until which we measured $C(t)$. 
We confirmed that $C(t_{\rm max})$ is almost zero (see Fig.~\ref{fig:Corr_3methods}~(i)). 
For small $\mu$, $\tau$ gradually increases with $\mu$ and it does not depend on the size. 
In contrast, $\tau$ for large $\mu$ depends not only on $\mu$ but also on $N_{\rm site}$, 
and it rapidly increases with them. This result is consistent with a previous result 
obtained by the cavity method that the model exhibits a dynamical 
transition with the breaking of ergodicity at $\mu_{\rm d}\approx 6.4$~\cite{Rivoire04}. 
This result indicates that, in the thermodynamic limit, the relaxation
time diverges at this chemical potential. 

\subsection{Comparison with the $N$-fold way method}
\label{subsec:Nfoldway}

In this subsection, we compare the LRMC method with the $N$-fold way method in detail. 
The following is the procedure of the conventional $N$-fold way method
without the particle-hole exchange update:
\begin{itemize}
\item[1)] Calculate the sum of insertion probabilities $P_{\rm ins}$ 
and that of deletion probabilities $P_{\rm del}$ in unit time by using insertion 
and deletion lists. If we employ the Metropolis transition probability, they are given as 
\begin{equation}
P_{\rm ins}\equiv K\{\sigma_i\},
\end{equation}
\begin{equation}
P_{\rm del}\equiv N\{\sigma_i\}{\rm e}^{-\mu}.
\end{equation}

\item[2)] Determine the residence time $\tau$ at the current state 
with exponential distribution
\begin{equation}
Q(\tau){\rm d}\tau =P_{\rm total}\exp(-P_{\rm total}\tau) {\rm d}\tau,
\end{equation}
where $P_{\rm total}\equiv P_{\rm ins}+P_{\rm del}$.

\item[3)] Increase the time by $\tau$. 

\item[4)] Determine an event which happens after the stay. 
The probability for insertion event and that for deletion event are 
$P_{\rm ins}/P_{\rm total}$ and $P_{\rm del}/P_{\rm total}$, respectively. 

\item[5)] If insertion is chosen in step 4), select an insertion site 
at random from the insertion list and insert a particle at the site. Otherwise, 
select a deletion site at random and delete a particle from the site.

\item[6)] Update the insertion and deletion lists. 

\item[7)] Return to 1).
\end{itemize}

An advantage of the $N$-fold way method over the insertion-deletion update in the 
LRMC method is that it is a rejection-free method. In the $N$-fold way method, 
the event determined in step 4) is always performed in step 5). We therefore 
expect that the efficiency of the $N$-fold way method is higher than that of 
the the insertion-deletion update in the LRMC method. Another characteristic 
of the $N$-fold way method is that residence times determined in step 2) 
differ from state to state. This means that each sampled state has a different weight. 
In contrast, each state sampled by the LRMC method has an equal weight. 
This property makes combinations between the LRMC method and other MC methods 
such as the replica exchange method and the Wang-Landau method simpler. 
However, it should be noted that such combinations are also possible in the 
$N$-fold way method~\cite{Schulz01}. 

\begin{figure}[t]
\begin{center}
\includegraphics[width=0.8\columnwidth]{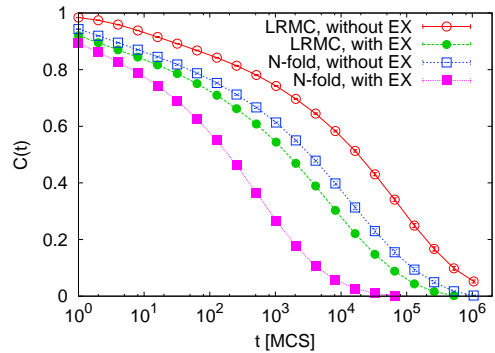}
\end{center}
\caption{(Color online) The time dependence of the autocorrelation functions 
measured with the LRMC method (circles) and the $N$-fold way method (squares). 
The data without the particle-hole exchange update and those with the particle-hole 
exchange update are denoted by open and full symbols, respectively. The model is the BM model 
on regular random graphs with $k=3$ and $l=1$. The number of sites $N_{\rm site}$ 
is $512$ and the chemical potential $\mu$ is $6.5$. The average over random graphs is taken 
over $100$ samples. For each sample, the average over thermal noise is taken over 
$320$ independent MC runs with different random number sequences.}
\label{fig:LRMC_vs_N-foldway}
\end{figure}

To evaluate the efficiency of the $N$-fold way method, 
we measured the autocorrelation function defined by Eq.~(\ref{eqn:DefAF}). 
In this calculation, we define one MC step 
by $N_{\rm site}$ updates in the particle configuration. 
The conditions of measurements are the same as those in the previous subsection. 
The results are shown in Fig.~\ref{fig:LRMC_vs_N-foldway}. 
The number of sites $N_{\rm site}$ is $512$ and the chemical potential $\mu$ 
is $6.5$. We also show the autocorrelation functions in the LRMC method, 
which have already been shown in Fig.~\ref{fig:Corr_3methods}, 
for comparison. The data without the particle-hole exchange update 
and those with the particle-hole exchange update are denoted by 
open and full symbols, respectively. As expected, $C(t)$'s of the $N$-fold way method 
decay faster than those of the LRMC method 
regardless of whether we use the particle-hole exchange update or not. 
Because the computational time of the $N$-fold way method per one MC step is 
comparable with that of the LRMC method, this result shows that the $N$-fold way 
method is superior to the LRMC method in efficiency. However, we have checked that 
the difference between the two methods in efficiency becomes smaller and smaller 
as the acceptance ratio of the insertion-deletion update increases. 
As shown in Fig.~\ref{fig:PacceptB}, the acceptance ratio increases 
with increasing $N_{\rm site}$ or decreasing $\mu$. 

\begin{figure}[t]
\begin{center}
\includegraphics[width=0.8\columnwidth]{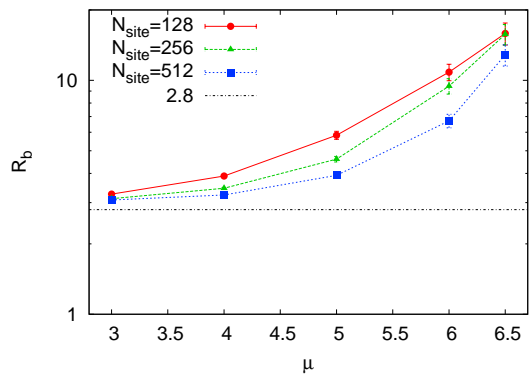}
\end{center}
\caption{(Color online) The $\mu$ dependence of the scaling factor $R_{\rm b}$ 
in the $N$-fold way method. The data for $N_{\rm site}=128$, $256$, and $512$ are 
denoted by full circles, full triangles, and full squares respectively. 
The model is the BM model on regular random graphs with $k=3$ and $l=1$.
The particle-hole exchange update is effective if $R_{\rm b}$ 
is larger than a threshold value $2.8$ depicted by the dash-dotted line (see text for details).
}
\label{fig:Rb_for_Nfoldway}
\end{figure}

We also see from Fig.~\ref{fig:LRMC_vs_N-foldway} that $C(t)$ of the LRMC method with 
the particle-hole exchange update decays slightly faster than that of the conventional 
$N$-fold way method without the particle-hole exchange update. Even if we take 
the difference in the computational time per one MC step into account, 
we can fairly say that they are comparable in efficiency.

Figure~\ref{fig:LRMC_vs_N-foldway} also shows that the decay of the autocorrelation 
function in the $N$-fold way method is accelerated by the particle-hole exchange update. 
To examine whether the particle-hole exchange update is effective or not in more detail, 
we performed an analysis similar to that in Fig.~\ref{fig:Rb}. The result is shown 
in Fig.~\ref{fig:Rb_for_Nfoldway}. Because the computational time of 
the $N$-fold way method per one MC step with the particle-hole exchange update 
is about $2.8$ times larger than that without the particle-hole exchange update, 
the particle-hole exchange update is effective if the scaling factor 
$R_{\rm b}$ is larger than $2.8$. We see that the efficiency of the $N$-fold way 
method is improved by the the particle-hole exchange update for all of the sizes 
and chemical potentials. In particular, it is rather effective when the chemical 
potential is large. 

\subsection{Effect of local update method on the replica exchange method}
\label{subsec:RexMC}

\begin{figure}[t]
\begin{center}
\includegraphics[width=0.8\columnwidth]{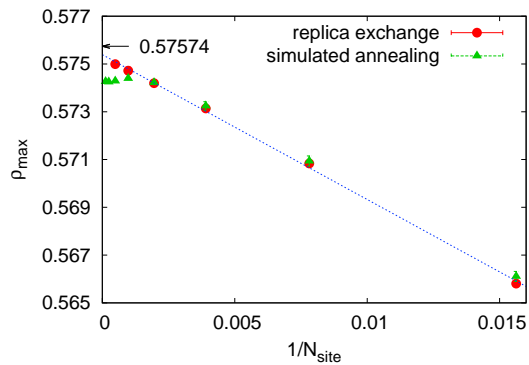}
\end{center}
\caption{(Color online) The sample averages of a maximum density 
$\rho_{\rm max}$ measured in two different simulation methods are compared. 
The data measured in the replica exchange method and those in the simulated 
annealing method are denoted by full circles and full triangles, respectively. In the figure, 
$\rho_{\rm max}$ is plotted as a function of $1/N_{\rm site}$. The straight line is a fitting 
line obtained from the data of the replica exchange method.}
\label{fig:close_packing}
\end{figure}

As we mentioned before, extended ensemble methods such as 
the multicanonical method~\cite{BergNeuhaus91,BergNeuhaus92}, 
the Wang-Landau method~\cite{WangLandau01A,WangLandau01B}, 
and the exchange MC method~\cite{HukushimaNemoto96} are 
known to be quite effective to relieve the problem of 
slow equilibration in glassy systems. 
The effectiveness of the extended ensemble methods is demonstrated in Fig.~\ref{fig:close_packing}. 
In the figure, the sample average of a maximum density $\rho_{\rm max}$ observed 
during simulation is plotted as a function of $1/N_{\rm site}$. 
The measurement is performed by both the replica exchange 
and simulated annealing methods. When $N_{\rm site}$ is small, there is no difference 
between the two data. However, we clearly see the difference for large $N_{\rm site}$. 
$\rho_{\rm max}$ measured by the simulated annealing method is saturated to a value 
around $0.5742$, whereas $\rho_{\rm max}$ measured by the replica exchange method 
continues to increase with the size. Furthermore, the extrapolated value of $\rho_{\rm max}$ 
in the thermodynamic limit agrees well with the close-packing density $\rho_{\rm s}=0.57574$ 
obtained by the cavity method~\cite{Rivoire04,Krzakala08}.

\begin{figure}[t]
\begin{center}
\includegraphics[width=0.8\columnwidth]{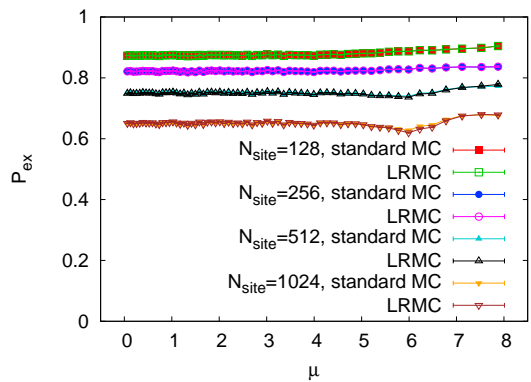}
\end{center}
\caption{(Color online) The chemical potential $\mu$ dependence of the acceptance 
ratio of the replica exchange $P_{\rm ex}$ for several sizes. 
The model is the BM model on regular random graphs with $k=3$ and $l=1$. 
The average over random graphs is taken over $100$ samples.
The data with the standard MC method and those with 
the LRMC method are denoted by full and open symbols, 
respectively. They completely collapse into each other.}
\label{fig:ProbEX}
\end{figure}

\begin{figure}[t]
(i)
\begin{center}
\includegraphics[width=0.8\columnwidth]{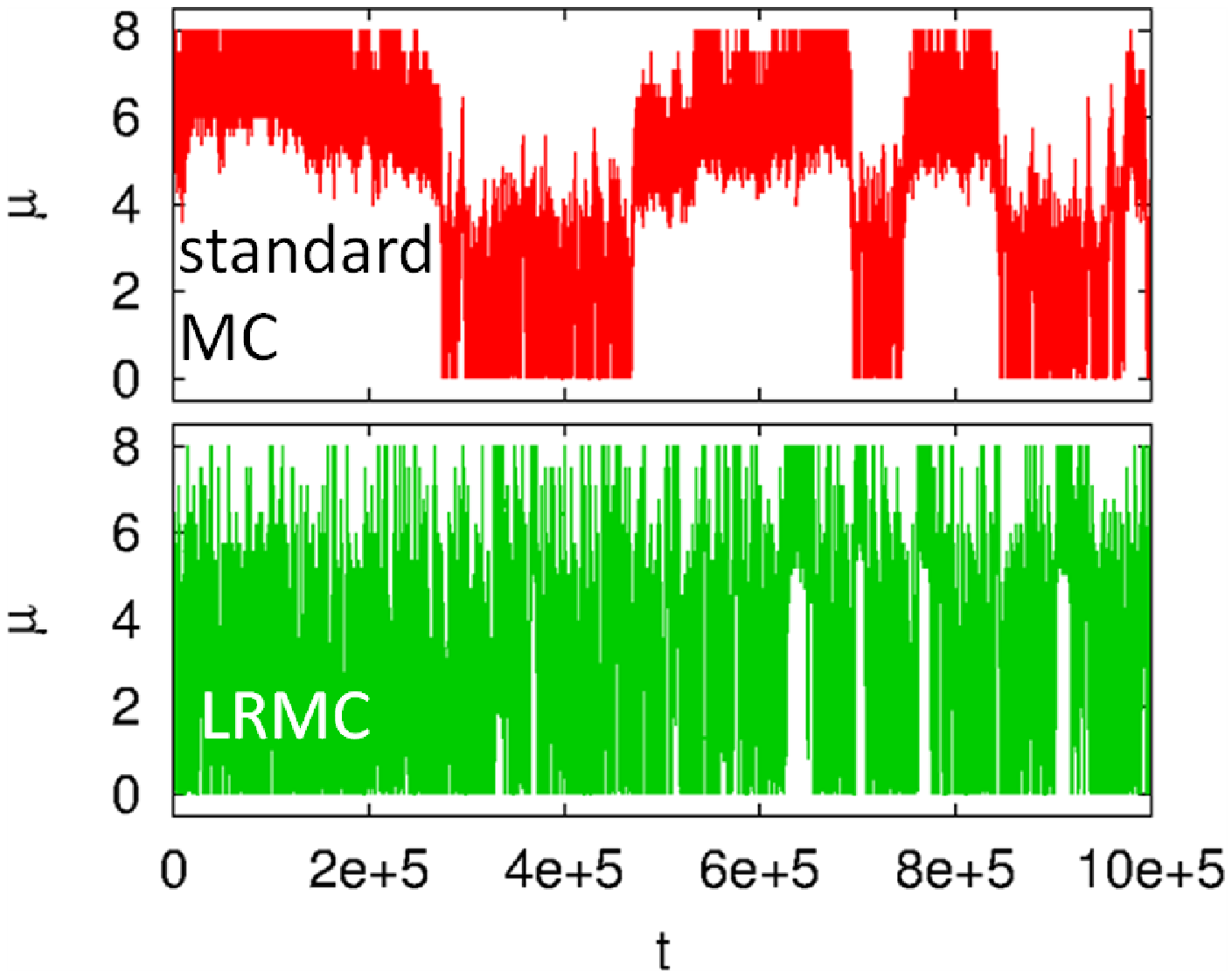}
\end{center}
(ii)
\begin{center}
\includegraphics[width=0.8\columnwidth]{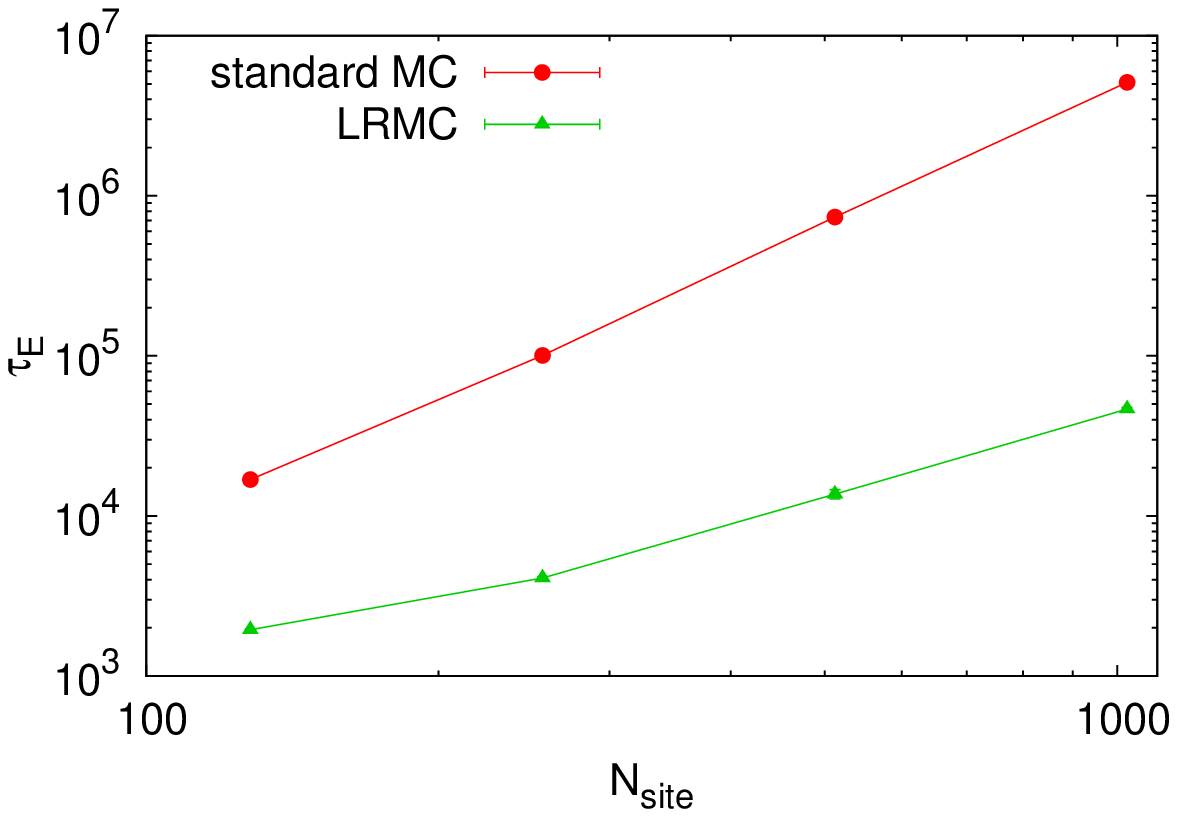}
\end{center}
\caption{(Color online) (i) The time dependences of the chemical potential 
of a specific replica measured with two different local update methods are compared. 
The data with the standard MC method and those with the LRMC method are shown 
in the upper and lower panels, respectively. The number of sites $N_{\rm site}$ is $512$. 
(ii) The size dependences of the ergodic time $\tau_{\rm E}$ 
measured with two different local update methods are compared. The data with the standard 
MC method and those with the LRMC method are denoted by full circles and full triangles, 
respectively. In both figures, the model is the BM model on regular random graphs 
with $k=3$ and $l=1$. In the measurements of $\tau_{\rm E}$, the average over 
random graphs is taken over $100$ samples. }
\label{fig:ReplicaMovement}
\end{figure}

Because extended ensemble methods work well even in glassy systems, 
one may consider that the choice of local update method is not so important as long as 
we use an extended ensemble method. However, there is a possibility 
that the efficiency of extended ensemble methods depends on 
the efficiency of local update methods. 
In fact, it has been pointed out that details of local update affect the efficiency of the replica 
exchange method~\cite{Bittner08}. We therefore examined how the efficiency 
of the replica exchange method concerning chemical potential 
depends on the choice of the local update method. The local update methods 
we examined are the standard MC method and the LRMC method 
(the method (c) in \S\ref{subsec:efficiency}). In the replica exchange method, 
every two replicas at adjacent chemical potentials $\mu_i$ and $\mu_{i+1}$ 
are attempted to be exchanged per one MC step. We set the lowest chemical potential 
and the highest one at $0$ and $8$, respectively. 
The intermediate chemical potentials between them are determined 
so that the acceptance ratio of the replica exchange $P_{\rm ex}$ 
is roughly constant. The number of replicas is $64$. A common set of 
chemical potentials is used for two simulations with different local update methods. 
To evaluate the efficiency of the replica exchange method, we measured the ergodic time 
$\tau_{\rm E}$. This is defined by the average MC steps for each replica to move from the 
highest chemical potential to the lowest one and return to the highest one. 
Because the system quickly forgets the current particle configuration 
at low $\mu$'s, the relaxation time of the replica exchange method 
is roughly estimated by the ergodic time.

In Fig.~\ref{fig:ProbEX}, the acceptance ratio of the replica exchange $P_{\rm ex}$ 
with the standard MC method and that with the LRMC method 
are plotted as a function of $\mu$. We see that the two acceptance ratios
are completely the same at all of the chemical potentials and sizes. 
We next show in Fig.~\ref{fig:ReplicaMovement}~(i) how the chemical potential of 
a specific replica changes with time by the replica exchange process. 
The number of sites $N_{\rm site}$ is $512$. We clearly see that 
the movement of the replica with the standard MC method is different 
from that with the LRMC method. In the former case, 
there is a bottle-neck in the replica movement around $\mu \approx 4$. 
In contrast, such bottle-neck does not exist in the latter case. 
Figure~\ref{fig:ReplicaMovement}~(ii) shows the size dependence of the ergodic time $\tau_{\rm E}$ 
of the two local update methods. As expected from Fig.~\ref{fig:ReplicaMovement}~(i), 
the ergodic time with the standard MC method is much larger than that with the LRMC method. 
The ratio of the former to the latter increases with the size and it reaches more than 
$10^2$ when $N_{\rm site}=1024$. These results show that the efficient 
local update method is important to make the replica exchange method more effective. 

\begin{figure}[t]
\begin{center}
\includegraphics[height=0.8\columnwidth,angle=270]{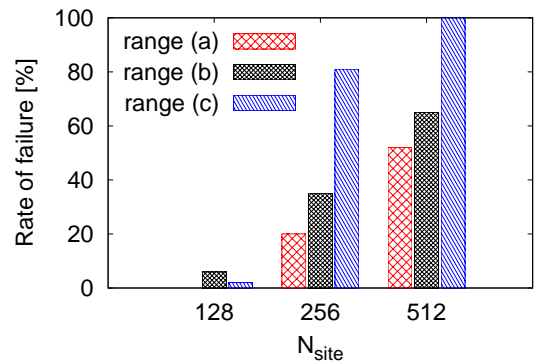}
\end{center}
\caption{(Color online) The rate of failure of the DOS measurement by the Wang-Landau method 
with the standard MC method. The model is the BM model on regular random graphs with $k=3$ 
and $l=1$. The DOS was measured for 100 samples with different random graphs 
at each $N_{\rm site}$ and the rate of failure was calculated from the results of these measurements.}
\label{fig:WLincomplete}
\end{figure}

\subsection{Effect of local update method on the Wang-Landau method}
\label{subsec:DOS}
In this subsection, we examine how the choice of the local update method 
affects the efficiency of the density of state (DOS) measurement by 
the Wang-Landau method~\cite{WangLandau01A,WangLandau01B}. 
In the BM model, the DOS is defined by 
\begin{equation}
\Omega(N') \equiv {\rm Tr}_{\{ \sigma_i \}} \delta (N'-N\{ \sigma_ i \})C\{ \sigma_i \}, 
\label{eqn:def_DOS}
\end{equation}
where $C\{\sigma_i\}$ is the indicator function which appears in Eq.~(\ref{eqn:distribution}). 

The local update methods we examined are the same as in the previous subsection, 
{\it i.e.}, the standard MC method and the LRMC method (the method (c) in \S\ref{subsec:efficiency}). 
The DOS is measured in the range of $0\le\rho\equiv N/N_{\rm site}\le 0.572$. 
The calculation of the DOS is parallelized by dividing the whole range 
into three sub-ranges and calculating the DOS in each sub-range independently. 
To be specific, we set the three ranges as follows: 
\begin{itemize}
\item[(a)] $0\le \rho \le 0.55$.
\item[(b)] $0.54 \le \rho \le 0.565$. 
\item[(c)] $0.56 \le \rho \le 0.572$. 
\end{itemize}
We set the lower bounds of the three ranges so that they are lower than the  dynamical  
transition density $\rho_{\rm d}=0.5708$~\cite{Rivoire04,Krzakala08} at which the ergodicity is broken. 
Because the three ranges include a region $\rho\le \rho_{\rm d}$ where a fast mixing is realized, 
we can expect that the Wang-Landau method efficiently calculate the DOS. 
By the Wang-Landau method, we can only calculate the relative DOS 
which is proportional to the absolute DOS defined by Eq.~(\ref{eqn:def_DOS}). 
We therefore calculate the absolute DOS by the following procedure: We first adjust the ratios 
of the proportionality constants of adjacent ranges to connect them as well as possible. 
It should be noted that there is overlap between two adjacent ranges. 
We next use the condition $\Omega(0)=1$ to determine the three proportionality constants. 
This condition comes from the fact that the number of particles is zero if and only if 
all of the sites are empty.

\begin{figure}[t]
\begin{center}
\includegraphics[width=0.8\columnwidth]{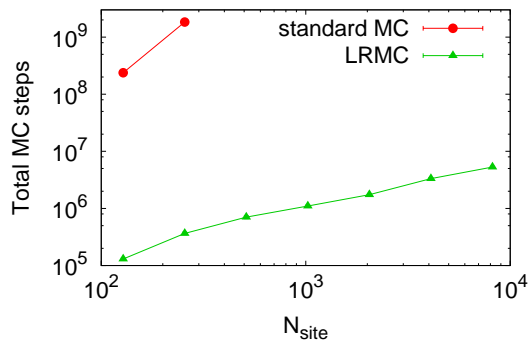}
\end{center}
\caption{(Color online) The size dependences of the total MC steps 
of the DOS calculations with two different local update methods are compared. 
The data with the standard MC method and those with the LRMC method are denoted 
by full circles and full triangles, respectively. The total MC steps are the sum of MC steps of 
all of the three ranges. The DOS was measured for 100 samples with different 
random graphs at each $N_{\rm site}$ and the average is taken over the samples for which 
we succeeded in calculating the DOS against all of the three ranges. The model is the BM model 
on regular random graphs with $k=3$ and $l=1$. }
\label{fig:Nsite_vs_WLtime}
\end{figure}

To measure the DOS, we used the standard procedure of the Wang-Landau method. 
The detailed conditions are as follows: 
We set the initial value of the modification factor $f$ to $\exp(0.1)$. 
The histogram is checked every $1000$ MC steps. 
We regard the histogram $H(N)$ as flat when $H(N)$ for all $N$ 
is not less than 80\% of their mean value. 
If the histogram is flat, we reduce the modification factor as 
$f_{k+1}=\sqrt{f_k}$ and reinitialize $H(N)$ to zero. 
We stop our simulation after we reduce the modification factor 20 times. 
Because we set the initial value of $f$ to $\exp(0.1)$, 
the final value of $f$ is $\exp(0.1\times 2^{-20})$. 
We checked that the calculated DOS converges well in later stages of simulations. 
The numbers of sites we examined were $2^m$ $(m=7,8,\cdots,13)$ for the calculations 
with the LRMC method, and $2^n$ $(n=7,8,9)$ for those with the standard MC method. 
For each $N_{\rm site}$, we calculated the DOS for $100$ samples 
with different random graphs. 
We set the maximum MC steps at each modification factor to $5\times 10^8$. 
If the histogram does not become flat until this MC steps, 
we regard the DOS calculation as failed and stop the simulation of the sample. 

Figure~\ref{fig:WLincomplete} shows the rate of failure of the DOS calculation 
with the standard MC method. The rate is measured for each range. 
We see that it increases with increasing $N_{\rm site}$. 
The rate of failure of the range (c) is $100\%$ when $N_{\rm site}$ is $512$. 
In contrast, when we used the LRMC method as a local update, the rates of failure 
were $0\%$ for all of $N_{\rm site}$'s and ranges. We also examined how the total MC steps of 
the DOS measurement, which are the sum of MC steps of all of the three ranges, 
depend on $N_{\rm site}$. In Fig.~\ref{fig:Nsite_vs_WLtime}, we plot the average 
of the total MC steps as a function of $N_{\rm site}$. The average is calculated from 
the samples for which we succeeded in calculating the DOS against all of the three ranges. 
We do not show the data at $N_{\rm site}=512$ for the standard MC method 
because the rate of failure of the range (c) was $100\%$. 
Because we exclude the samples in which we failed to measure the DOS, 
the average of total MC steps is underestimated in the standard MC method. 
Nevertheless, the average of total MC steps in the standard MC method is 
more than $1000$ times larger than that in the LRMC method even at $N_{\rm site}=128$. 
This result shows that the efficiency of the Wang-Landau method is much improved by 
the use of the LRMC method.

\begin{figure}[t]
\begin{center}
\includegraphics[width=0.8\columnwidth]{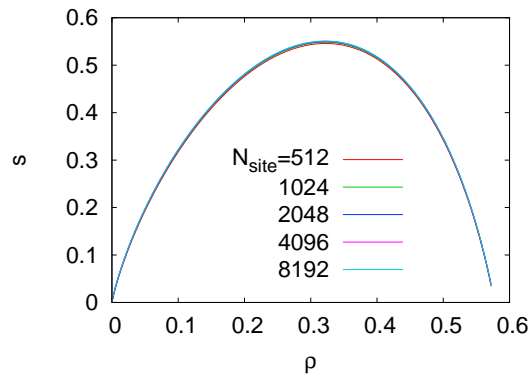}
\end{center}
\caption{(Color online) The density $\rho$ dependence of the entropy per site for five sizes. 
The entropy is measured by the Wang-Landau method with the LRMC method. 
The model is the BM model on regular random graphs with $k=3$ and $l=1$. 
The average over random graphs is taken over $100$ samples.}
\label{fig:DOS}
\end{figure}

In Fig.~\ref{fig:DOS}, the entropy per site $s(N) = S(N)/N_{\rm site}$ 
is plotted as a function of $\rho$ for several $N_{\rm site}$'s, 
where the entropy is defined by
\begin{equation}
S(N)\equiv \overline{\log\{\Omega(N)\}}.
\end{equation}
We only show the data with the LRMC method in the figure. 
We see that all of the data almost completely collapse into 
a single curve. $s(\rho)$ starts to drop around $\rho\approx 0.32$ and it becomes 
close to zero at the highest density $\rho=0.572$ of our calculation. 
This result is reasonable because the close-packing density $\rho_{\rm s}$ 
is estimated to be $0.57574$ by the cavity method~\cite{Rivoire04,Krzakala08}.

\section{Conclusions}
\label{sec:conclusions}
In this paper, we have presented an efficient Monte Carlo method called the LRMC method 
for lattice glass models which are characterized by hard constraint conditions. 
Like the $N$-fold way method, we make a list of the sites into which we can insert a particle, 
and update it whenever the particle configuration is changed. By utilizing the list, 
we can avoid a useless trial of insertion which conflicts with the constraint conditions. 
The efficiency of the LRMC method with the particle-hole exchange update
is much higher than that of the standard Monte-Carlo method, 
and it is comparable with that of the conventional $N$-fold way method without the particle-hole 
exchange update. We also found that the particle-hole exchange update is rather effective 
for the $N$-fold way method. The efficiency of the $N$-fold way method is considerably improved 
by adopting the particle-hole exchange update into the method. We also have shown 
that the efficiency of the replica exchange method and that of the Wang-Landau method are improved 
much by using the LRMC method as a local update method. This result shows that the efficient local 
update method is quite important to make these extended ensemble methods more effective.

In the present study, the LRMC method was applied only to the BM model on a regular random graph. 
However, like the $N$-fold way method, the LRMC method is applicable no matter whether the model 
is defined on a sparse random graph or a usual lattice in finite dimensions. 
Furthermore, an applicable class of models includes not only general lattice glass models 
but also constraint-satisfaction problems with binary variables such as $K$-satisfiability 
problems and vertex cover problems~\cite{MezardMontanariBook}. We hope that our study will 
stimulate further research in this field. 

\section*{Acknowledgment}
This work is supported by Grant-in-Aids for Scientific Research (No.~22340109 and No.~25400387) 
from the Ministry of Education, Culture, Sports, Science and Technology in Japan.


\appendix

\section{The deviation of Eq.~(\ref{eqn:KNrelation_final})}
\label{sec:NewAppendix}

\begin{figure}[t]
\begin{center}
\includegraphics[width=0.8\columnwidth]{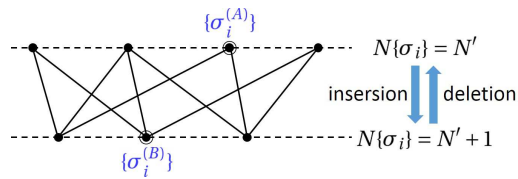}
\end{center}
\caption{(Color online) Schematic illustration of transitions between 
states with $N\{\sigma_i \}=N'$ and those with $N\{\sigma_i \}=N'+1$. 
The solid lines denote transitions between states. In the figure, 
$K\{\sigma_i^{(A)}\}=2$ and $N\{\sigma_i^{(B)}\}=3$. 
The total number of transitions $N_{\rm trans}$ is calculated to be $9$ 
by summing up either $N\{\sigma_i\}$'s of the states with $N\{\sigma_i \}=N'+1$ 
or $K\{\sigma_i\}$'s of the states with $N\{\sigma_i \}=N'$. 
}
\label{fig:Ntrans}
\end{figure}

We now consider to calculate the total number of transitions $N_{\rm trans}$ 
between states with $N\{\sigma_i \}=N'$ and those with $N\{\sigma_i \}=N'+1$. 
As shown in Fig.~\ref{fig:Ntrans}, 
we can estimate $N_{\rm trans}$ by counting either transitions 
caused by deleting a particle or those caused by inserting a particle. 
It is obvious that the two estimations give the same number. 
From the former estimation, we obtain
\begin{equation}
N_{\rm trans}=\mathrm{Tr}^{(N'+1)} N\{\sigma_i\},
\label{eqn:transA}
\end{equation}
where $\mathrm{Tr}^{(N')}$ denotes the sum over particle configurations with $N\{\sigma_i\}=N'$. 
In Eq.~(\ref{eqn:transA}), we have used the fact that 
the system can translate from a state $\{ \sigma_i \}$ to 
$N\{ \sigma_i \}$ different states because deletion of a particle 
never conflicts with the constraint conditions. On the other hand, 
we obtain from the latter estimation that
\begin{equation}
N_{\rm trans}=\mathrm{Tr}^{(N')} K\{\sigma_i\}.
\label{eqn:transB}
\end{equation}
From Eqs.~(\ref{eqn:transA}) and (\ref{eqn:transB}), we obtain
\begin{equation}
\mathrm{Tr}^{(N'+1)} N\{\sigma_i\}=\mathrm{Tr}^{(N')} K\{\sigma_i\}.
\label{eqn:KNrelation}
\end{equation}
This equation is valid for $0\le N' \le N_{\rm max}-1$, 
where $N_{\rm max}$ is the maximum number of particles among all of possible 
particle configurations. 

The average number of particles $\langle N\{ \sigma_i \} \rangle_{\mu}$ in the 
ground-canonical ensemble is given as 
\begin{equation}
\langle N\{ \sigma_i \} \rangle_{\mu}=Z_{\rm G}^{-1}\sum_{N'=1}^{N_{\rm max}}
\exp[\mu N']\mathrm{Tr}^{(N')} N\{\sigma_i\},
\label{eqn:def_Nave}
\end{equation}
where $Z_{\rm G}$ is the grand partition function. Note that the lower bound of 
the sum in the right hand side is one. In a similar way, 
$\langle K\{ \sigma_i \} \rangle_{\mu}$ is given as 
\begin{equation}
\hspace*{-1mm}\langle K\{ \sigma_i \} \rangle_{\mu}=Z_{\rm G}^{-1}\sum_{N'=0}^{N_{\rm max}-1}
\exp[\mu N']\mathrm{Tr}^{(N')} K\{\sigma_i\}.
\label{eqn:def_Kave}
\end{equation}
We set the upper bound of the sum to $N_{\rm max}-1$ because 
$K\{\sigma_i\}=0$ for states with the maximum number of particles. 
By substituting Eq.~(\ref{eqn:KNrelation}) into Eq.~(\ref{eqn:def_Kave}), 
we obtain
\begin{align}
&\langle K\{ \sigma_i \} \rangle_{\mu} \nonumber\\
& =Z_{\rm G}^{-1}\sum_{N'=0}^{N_{\rm max}-1}\exp[\mu N']\mathrm{Tr}^{(N'+1)} 
N\{\sigma_i\} \nonumber\\
& ={\rm e}^{-\mu}Z_{\rm G}^{-1}\sum_{N'=0}^{N_{\rm max}-1}\exp[\mu (N'+1)]\mathrm{Tr}^{(N'+1)} 
N\{\sigma_i\}
\nonumber \\
& = {\rm e}^{-\mu}\langle N\{ \sigma_i \} \rangle_{\mu},
\end{align}
where we have used Eq.~(\ref{eqn:def_Nave}) to go from the third line to the fourth, 
eventually Eq.~(\ref{eqn:KNrelation_final}) is obtained.

\section{The method to detect the sites on which the insertion list has to be updated}
\label{sec:appendixA}
In the LRMC method, we need to update the insertion list 
whenever a particle is inserted into a site or deleted from there. 
In order to do that, we first need to detect the sites on which the insertion 
list has to be updated. In this appendix, we describe the method to detect them. 
We emphasize that it is quite important to perform this procedure as efficient 
as possible because this is one of the most fundamental procedure in the LRMC method. 

We introduce two variables to explain the method. 
Firstly, a Boolean ${\tt Insert}(p)$ is ${\tt .TRUE.}$ if it is possible 
to insert a particle at the site $p$, or it is ${\tt .FALSE.}$ otherwise. 
The purpose of this appendix is to detect the sites at which ${\tt Insert}(p)$ 
changes from ${\tt .TRUE.}$ to ${\tt .FALSE.}$ or conversely by insertion 
or deletion of a particle. Secondly, an integer ${\tt NNN}(p)$ denotes 
the number of the nearest-neighbouring particles at site $p$. 
Due to the constraint conditions of the BM model, ${\tt NNN}(p)$ has to be less than or equal to $l$ 
if the site $p$ is occupied by a particle. After we choose an initial particle configuration, 
we set ${\tt Insert}(p)$ and ${\tt NNN}(p)$ at the beginning of the simulation. 
Then, we update them whenever a particle is inserted or deleted.

The organization of this appendix is as follows: In \S\ref{subsec:A1}, we explain basic strategy 
to detect the sites at which ${\tt Insert}(p)$ changes by insertion or deletion of a particle. 
In this subsection, we consider the general case that the two parameters $k$ and $l$ of the BM model 
are arbitrary integers. In \S\ref{subsec:A2}, we consider a special case $l=1$ and explain 
an optimized method for this case. In these two subsections, 
we assume that the graph does not have a loop which involves an insertion or deletion site. 
We consider in \S\ref{subsec:A3} how the method to detect the changed sites should be modified 
when such loop exists.

\begin{figure}[t]
\begin{center}
\includegraphics[width=0.8\columnwidth]{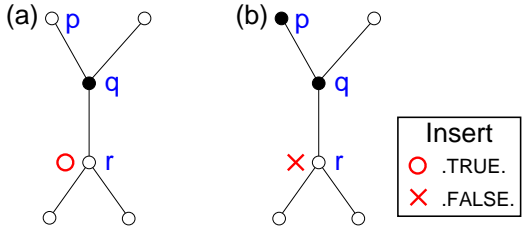}
\end{center}
\caption{(Color online) An example when ${\tt Insert}(r)$ depends on $\sigma_p$, 
where $r$ is a next nearest-neighbouring site of $p$. The values of the two parameters of the BM model 
are $k=3$ and $l=1$. In the particle configuration (b), it is impossible to insert a particle 
into the site $r$ because the number of the nearest-neighbouring particles of the site $q$ becomes 
larger than $l=1$ by the insertion. 
}
\label{fig:A1}
\end{figure}

\subsection{Basic strategy in the general BM model}
\label{subsec:A1}
In this subsection, we consider the general case that the two parameters $k$ and $l$ of the BM model 
are arbitrary integers, and explain the basic strategy to detect the sites at which ${\tt Insert}$ changes 
from ${\tt .TRUE.}$ to ${\tt .FALSE.}$ or conversely by insertion or deletion of a particle. 
Firstly, in the following subsection, we consider at which sites such change in ${\tt Insert}$ 
may occur. It is important to reduce the possible sites as many as possible to reduce the computational 
time. Secondly, in \S\ref{subsubsec:A12}, we explain the general method to judge whether 
a change in ${\tt Insert}$ really occurs or not.

\subsubsection{Specification of the possible sites}
\label{subsubsec:A11}

We consider to insert or delete a particle at a site $p$. 
Then, $\sigma_p$ changes from $0$ to $1$ or conversely. 
There are three kinds of sites where ${\tt Insert}$ 
may change: the site $p$ itself, the nearest-neighbouring sites of $p$, 
and the next nearest-neighbouring sites of $p$. 
Firstly, let us consider the case of the site $p$ itself. 
After we delete a particle from the site $p$, ${\tt Insert}(p)$ becomes ${\tt .TRUE.}$. 
On the other hand, after we insert a particle into the site $p$, 
${\tt Insert}(p)$ becomes ${\tt .FALSE.}$ because the site $p$ has already 
been occupied by a particle. These two facts mean that ${\tt Insert}(p)$ always changes 
by the insertion or deletion of a particle at the site $p$. 
Therefore, we do not need to judge whether the change occurs or not. 
Secondly, concerning a nearest-neighbouring site $q$, 
it is apparent that ${\tt Insert}(q)$ is ${\tt .FALSE.}$ if the site 
$q$ is already occupied by a particle. 
This means that we do not need to judge 
whether ${\tt Insert}(q)$ changes or not if $\sigma_q$ is $1$. 
Lastly, we consider the case of a next nearest-neighbouring site $r$. 
As it is shown in Fig.~\ref{fig:A1}, if ${\tt Insert}(r)$ depends on $\sigma_p$, 
the nearest-neighbouring site $q$ which is between $p$ and $r$ has to be occupied by a particle. 
In other words, we do not need to judge whether ${\tt Insert}(r)$ changes or not 
if $\sigma_q$ is $0$. 

By taking all of these considerations into account, we find that the following procedure 
is enough to list all of the sites where ${\tt Insert}$ changes by the insertion 
or deletion of a particle at the site $p$:
\begin{itemize}
\item[(I)] Add the site $p$ into the list. 
\item[(II)] Perform the following procedure for all of the nearest-neighbouring sites $q$:
\begin{itemize}
\item[(i)] If $\sigma_q=0$, judge whether ${\tt Insert}(q)$ changes or not. 
If it changes, add the site $q$ into the list. 
\item[(ii)] If $\sigma_q=1$, check all of the next nearest-neighbouring sites $r$ 
which connect with the site $q$ whether ${\tt Insert}(r)$ changes or not. 
If it changes, add the site $r$ into the list. 
\end{itemize}
\end{itemize}

\begin{figure}[t]
\begin{center}
\includegraphics[width=1.0\columnwidth]{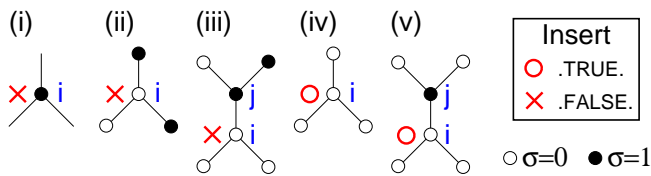}
\end{center}
\caption{(Color online) Three examples when one of the three necessary and sufficient conditions 
for ${\tt Insert}(i)={\tt .TRUE.}$ is not satisfied, and two examples when all of the three 
conditions are satisfied. The values of the two parameters of the BM model are $k=3$ and $l=1$. 
The first, the second, and the third condition
are not satisfied in the case (i), (ii), and (iii), respectively. 
In contrast, all of the three conditions are satisfied in the cases (iv) and (v). 
}
\label{fig:A2}
\end{figure}

\subsubsection{General method to judge whether {\tt Insert} changes or not}
\label{subsubsec:A12}
When ${\tt Insert}(q)$ may change at a site $q$ by the insertion or deletion of a particle 
at the site $p$, we have to judge whether the change really occurs or not. 
In this subsection, we consider how we judge it. Because, as mentioned above, 
${\tt Insert}$ is updated whenever a particle is inserted or deleted, 
we know whether ${\tt Insert}(q)$ is ${\tt .TRUE.}$ or ${\tt .FALSE.}$ 
before the insertion or deletion. Therefore, to judge whether the change occurs or not, 
it is sufficient to know whether ${\tt Insert}(q)$ is ${\tt .TRUE.}$ or ${\tt .FALSE.}$ 
{\it after} the insertion or deletion. As shown in Fig.~\ref{fig:A2}, 
the necessary and sufficient conditions for ${\tt Insert}(i)={\tt .TRUE.}$ 
are as follows:
\begin{itemize}
\item $\sigma_i=0.$
\item ${\tt NNN}(i)\le l.$
\item ${\tt NNN}(j)\le l-1$ for all of the nearest-neighbouring sites $j$ of the site $i$ 
which are occupied by a particle. 
\end{itemize}
Therefore, ${\tt Insert}(i)$ is ${\tt .TRUE.}$ if all of the three conditions are 
satisfied. Otherwise, it is ${\tt .FALSE.}$.

\subsection{Optimized method for $l=1$}
\label{subsec:A2}
In principle, we can judge whether ${\tt Insert}$ changes or not by the method described 
in \S\ref{subsubsec:A12}. However, it is time-consuming to examine whether all of the three 
necessary and sufficient conditions mentioned in \S\ref{subsubsec:A12} are satisfied or not 
for all of the sites at which ${\tt Insert}$ may change. In this subsection, we consider 
a special case $l=1$, and consider to reduce the procedure for the judgement 
as much as possible. 

Before we describe the details of the method, for the convenience of explanation, 
we show the necessary and sufficient conditions for ${\tt Insert}(i)={\tt .TRUE.}$ 
in the case of $l=1$:
\begin{itemize}
\item[(a)] $\sigma_i=0$.
\item[(b)] ${\tt NNN}(i)$ is either $0$ or $1$. 
\item[(c)] If ${\tt NNN}(i)$ is $1$, ${\tt NNN}(j)=0$, where $j$ is the nearest-neighbouring site 
which is occupied by a particle. 
\end{itemize}

\subsubsection{Judgement at a next nearest-neighbouring site}
\label{subsubsec:A21}

\begin{figure}[t]
\begin{center}
\includegraphics[width=1.0\columnwidth]{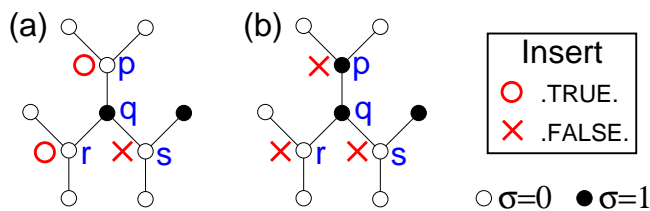}
\end{center}
\caption{(Color online) An example of particle configurations when we judge 
next nearest-neighbouring sites. 
The values of the two parameters of the BM model are $k=3$ and $l=1$. The particle configuration 
changes from (a) to (b) or conversely by inserting or deleting a particle at the site $p$, respectively. 
${\tt Insert(r)}$ changes by the insertion or deletion of a particle at the site $p$, 
whereas such change does not occur at the site $s$. }
\label{fig:A3}
\end{figure}

In this subsection, we consider to judge whether ${\tt Insert}(r)$ changes or not 
at a next nearest-neighbouring site $r$. Figure~\ref{fig:A3} shows an example of possible 
particle configurations when we judge the site $r$. 
The particle configuration changes from (a) to (b) or conversely by inserting or deleting 
a particle at the site $p$, respectively. 
We see that these particle configurations satisfy the following three conditions:
\begin{itemize}
\item[(1)] $\sigma_q=1$, where $q$ is the nearest-neighbouring site between $p$ and $r$. 
\item[(2)] Both $\sigma_r$ and $\sigma_s$ is $0$, where $r$ and $s$ are the next nearest-neighbouring 
sites which connect with $q$. 
\item[(3)] Both ${\tt Insert}(r)$ and ${\tt Insert}(s)$ are ${\tt .FALSE.}$ when $\sigma_p=1$. 
\end{itemize}
The first condition comes from the fact that we need to judge the site $r$ only if $\sigma_q$ is $1$ 
(see the procedure (ii) in the last paragraph of \S\ref{subsubsec:A11}). 
The second condition comes from the fact that the constraint condition 
has to be satisfied at the site $q$ even if $\sigma_p=1$. 
The third condition comes from the fact that ${\tt NNN}(q)$ has already been $1(=l)$ 
when $\sigma_p=1$. From the condition (3), 
we find that ${\tt Insert}(r)$ changes by inserting or deleting a particle 
at the site $p$ if and only if ${\tt Insert}(r)$ is ${\tt .TRUE.}$ when $\sigma_p=0$. 

Therefore, we next consider the necessary and sufficient conditions so that 
${\tt Insert}(r)$ is ${\tt .TRUE.}$ under the prior condition $\sigma_p=0$. 
As we mentioned above, there are the three necessary and sufficient conditions 
(a)-(c) mentioned at the beginning of of \S\ref{subsec:A2} 
so that ${\tt Insert}(r)$ is ${\tt .TRUE.}$. 
However, in practice, we do not need to check all of the three conditions. 
Firstly, the condition (a) is satisfied automatically due to the condition (2). 
Secondly, if ${\tt NNN}(r)$ is $1$ and the condition (b) is satisfied 
(note that it is impossible that ${\tt NNN}(r) = 0$ because $\sigma_q=1$), 
the condition (c), {\it i.e.}, ${\tt NNN}(q)=0$, is also satisfied automatically 
because $\sigma_p=0$ by the prior condition and 
both $\sigma_r$ and $\sigma_s$ are 0 by the condition (2). From these two facts, 
we find that the condition (b) is the only necessary and sufficient condition we have to check. 

In conclusion, ${\tt Insert}(r)$ changes by the insertion or deletion of a particle at the site $p$ 
if and only if ${\tt NNN}(r)$ is $1$. 

\subsubsection{Judgement at a nearest-neighbouring site after the insertion of a particle into the site $p$}
\label{subsubsec:A22}
In this subsection, we consider to judge whether ${\tt Insert}(q)$ changes or not 
by the insertion of a particle at the site $p$, where $q$ is a nearest-neighbouring site. 
Because we insert a particle into the site $p$, it is impossible that ${\tt Insert}(q)$ changes 
from ${\tt .FALSE.}$ to ${\tt .TRUE.}$ by the insertion. 
Therefore, the necessary and sufficient conditions that ${\tt Insert}(q)$ changes 
are that ${\tt Insert}(q)$ {\it before} the insertion is ${\tt .TRUE.}$ 
and that ${\tt Insert}(q)$ {\it after} the insertion is ${\tt .FALSE.}$. 
Because ${\tt Insert}$ is updated at each step, we know ${\tt Insert}(q)$ before the insertion. 
Therefore, we can easily check whether the first condition is satisfied or not. 
We next consider the second condition. To this end, it is convenient to consider 
the necessary and sufficient conditions for the complementary event that 
${\tt Insert}(q)$ after the insertion is ${\tt .TRUE.}$. 
In general, the necessary and sufficient conditions are the three conditions (a)-(c)
mentioned at the beginning of \S\ref{subsec:A2}. However, 
because we need to judge the site $q$ only if $\sigma_q$ is $0$ 
(see the procedure (i) in the last paragraph of \S\ref{subsubsec:A11}), 
the condition (a) is always satisfied. We also notice that ${\tt NNN}(q)$ can not 
be $0$ after the insertion at the site $p$. Therefore, 
the necessary and sufficient conditions that ${\tt Insert}(q)$ after the insertion is 
${\tt .TRUE.}$ are that ${\tt NNN}(q)=1$ and ${\tt NNN}(p)=0$. This means that 
${\tt Insert}(q)$ after the insertion is ${\tt .FALSE.}$ if and only if 
either ${\tt NNN}(q)\ne 1$ or ${\tt NNN}(p)\ne 0$, or both. 

In conclusion, ${\tt Insert}(q)$ changes from ${\tt .TRUE.}$ to ${\tt .FALSE.}$ 
if and only if the following two conditions are satisfied: 
\begin{itemize}
\item ${\tt Insert}(q)$ is ${\tt .TRUE.}$ before a particle is inserted at the site $p$. 
\item Either ${\tt NNN}(q)\ne 1$ or ${\tt NNN}(p)\ne 0$, or both, after a particle is inserted 
into the site $p$. 
\end{itemize}
It should be noted that ${\tt NNN}(q)$ and ${\tt NNN}(p)$ are the numbers of the nearest-neighbouring particles 
{\it after} a particle is inserted into the site $p$.

\subsubsection{Judgement at a nearest-neighbouring site after the deletion of a particle from the site $p$}
\label{subsubsec:A23}
In this subsection, we consider to judge whether ${\tt Insert}(q)$ changes or not 
by the deletion of a particle at the site $p$, where $q$ is a nearest-neighbouring site. 
From a consideration similar to that in the previous subsection, we find that 
the necessary and sufficient conditions that ${\tt Insert}(q)$ changes 
are that ${\tt Insert}(q)$ {\it before} the deletion is ${\tt .FALSE.}$ 
and that ${\tt Insert}(q)$ {\it after} the deletion is ${\tt .TRUE.}$. 
Firstly, as mentioned in the previous subsection, we can easily check 
whether the first condition is satisfied or not. Secondly, 
${\tt Insert}(q)$ {\it after} the deletion is ${\tt .TRUE.}$ if and only if 
all of the three conditions (a)-(c) is satisfied. However, because the site $q$ is judged 
only if $\sigma_q$ is $0$ (see the procedure (i) in the last paragraph of \S\ref{subsubsec:A11}), 
we can remove the condition (a) from them.

In conclusion, ${\tt Insert}(q)$ changes from ${\tt .FALSE.}$ to ${\tt .TRUE.}$ if and only if 
the following three conditions are satisfied: 
\begin{itemize}
\item ${\tt Insert}(q)$ is ${\tt .FALSE.}$ before a particle is deleted from the site $p$. 
\item ${\tt NNN}(q)$ is 0 or 1. 
\item If ${\tt NNN}(q)$ is 1, ${\tt NNN}(j)$ is $0$, where $j$ is the nearest-neighbouring site 
of $q$ which is occupied by a particle. 
\end{itemize}
We again emphasize that ${\tt NNN}(q)$ and ${\tt NNN}(j)$ are the numbers of the nearest-neighbouring 
particles {\it after} a particle is deleted from the site $p$.

\begin{figure}[t]
\begin{center}
\includegraphics[width=1.0\columnwidth]{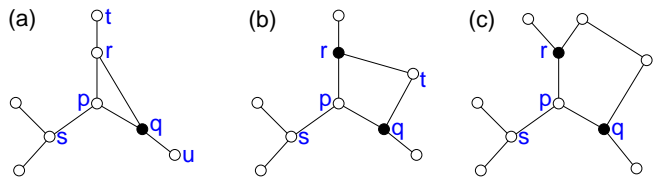}
\end{center}
\caption{(Color online) Three examples of a graph when a loop which involves the insertion or deletion site $p$ exists. 
The length of the loop is $3$ in the case (a), $4$ in the case (b), and $5$ in the case (c). 
}
\label{fig:A4}
\end{figure}

\subsection{The case when a loop exists around the insertion or deletion site}
\label{subsec:A3}

In \S\ref{subsec:A1} and \S\ref{subsec:A2}, we have implicitly assumed that there is no loop 
which involves the insertion or deletion site $p$. In this subsection, we explain
how we should change the method described in \S\ref{subsec:A1} and \S\ref{subsec:A2} 
when a loop which involves the site $p$ exists. Figure~\ref{fig:A4} shows examples of a graph 
when there is a loop. The length of the loop is $3$ in the case (a), $4$ in the case (b), 
and $5$ in the case (c). We can check that the method to judge whether ${\tt Insert}$ 
changes or not is still valid even if such loop exists. Therefore, we do not need to 
modify both the general method described in \S\ref{subsubsec:A12} and the optimized method 
for $l=1$ described in \S\ref{subsec:A2}. The only part we should change is 
the specification of the possible sites described in \S\ref{subsubsec:A11}. 

When a loop whose length is either $3$ or $4$ exists, {\it i.e.,} in the cases (a) and (b) in 
Fig.~\ref{fig:A4}, we should be careful not to check a site twice as a possible site. 
When the length of the loop is equal to or larger than $5$ like the case (c), 
we do not need to care about it. We hereafter consider the two cases (a) and (b). 
Firstly, when we check the site $q$ in the case (a), 
we should exclude the site $r$ from the next nearest-neighbouring sites so as not 
to check this site in the procedure (ii) in the last paragraph of \S\ref{subsubsec:A11}. 
Otherwise, the site $r$ is checked twice because this site is checked 
in the procedure (i) as a nearest-neighbouring site. To avoid this double check, 
in the procedure (ii), we should check a site $r$ when it satisfies 
the following two conditions:
\begin{itemize}
\item $r$ is a nearest-neighbouring site of $q$. 
\item $r$ is neither $p$ nor one of the nearest-neighbouring sites of $p$. 
\end{itemize}
Because these two conditions only depend on the shape of the graph, 
it is enough to calculate the set of sites which satisfy the two conditions 
once at the beginning of the simulation. This calculation should be done for each site. 
We next consider the case (b) in Fig.~\ref{fig:A4}. We suppose that, as shown in the figure, 
both $q$ and $r$ are occupied by a particle. Then, if we naively perform the procedure 
described in \S\ref{subsubsec:A11}, the site $t$ is checked twice as a possible site. 
When $l=1$, this kind of double check does not occur even if we naively perform the procedure 
because it is impossible that both $q$ and $r$ are occupied by a particle. 
Note that ${\tt NNN}(p)\le l =1$ because ${\tt Insert}(p)$ is ${\tt .TRUE.}$ 
when $\sigma_p=0$. However, if $l\ge 2$, we should modify the procedure 
so as to check the site $t$ {\it once} when both $q$ and $r$ are occupied by a particle.

\section{The method to update the insertion list}
\label{sec:appendixB}
In this appendix, we explain the method to update the insertion list. 
Before we start the explanation, we explain the situation when we update the insertion list 
and introduce several technical terms. We assume that we know all of the sites into which 
we can insert a particle. This information is stored in an array ${\tt List}$. 
The value of ${\tt List}(m)$ denotes the $m$-th site into which we can insert a particle. 
The number of the insertable sites is stored in an integer ${\tt Nlist}$. 
We assume that ${\tt List}(m)$ is $0$ for $m> {\tt Nlist}$. 
We also assume that we know where an insertable site $i$ is recorded 
in the array {\tt List}. This information is 
stored in an array ${\tt Reverse}(i)$ and the value of ${\tt Reverse}(i)$ 
denotes the position in the array ${\tt List}$. 
That is to say, if the number of ${\tt Reverse}(i)$ is, say, $10$, 
the site $i$ is insertable and ${\tt List}(10)=i$. 
It is worth noticing that, if we set ${\tt Reverse}(i)$ to a negative integer such as $-1$ 
when the site $i$ is not insertable, we do not need to prepare 
the array ${\tt Insert}(i)$ introduced in the appendix~\ref{sec:appendixA} 
because we can store the information whether the site $i$ is insertable or not 
in ${\tt Reverse}(i)$. In this setting, positive ${\tt Reverse}(i)$ corresponds to 
${\tt Insert}(i)={\tt .TRUE.}$ and negative ${\tt Reverse}(i)$ corresponds to 
${\tt Insert}(i)={\tt .FALSE.}$. After we choose an initial particle configuration, 
we set the two arrays ${\tt List}(m)$ and ${\tt Reverse}(i)$ and the integer ${\tt Nlist}$ 
at the beginning of the simulation. Then, we update them whenever a particle is 
inserted or deleted. 

We now start to explain the method to update the insertion list. We assume that 
we have detected all of the sites on which the insertion list has to be updated 
by using the method described in the appendix~\ref{sec:appendixA}. 
There are two operations to update the insertion list: 
\begin{itemize}
\item[(A)] Remove a site $i$ from the list. 
\item[(B)] Add a site $i$ into the list. 
\end{itemize}
We first consider the operation (A). Now the point is that, when we insert a particle, 
we choose the insertion site at random from the array ${\tt List}$. 
This means that the order in the list 
is not important. Therefore, when we remove the $m$-th element from the list 
($m={\tt Reverse}(i)$), 
we can replace the $m$-th element with the last element and set the last element 
to zero after that. 
By taking this and the fact that ${\tt Reverse}(i)=m$ into account, 
we can perform the operation (A) in the following way: 
\begin{verbatim}
 Reverse(List(Nlist))=Reverse(i)
 List(Reverse(i))=List(Nlist)
 List(Nlist)=0
 Nlist=Nlist-1
 Reverse(i)=-1
\end{verbatim}
The operation (B) is simpler. We just add the site $i$ at the end of the list. 
This is performed in the following way:
\begin{verbatim}
  Nlist=Nlist+1
  List(Nlist)=i
  Reverse(i)=Nlist
\end{verbatim}

\end{document}